\documentclass[twocolumn]{aastex62}
\pdfoutput=1
\usepackage{graphicx,rotating,amssymb,amsmath}
\usepackage{natbib}
\usepackage{multirow}

\newcommand\kms{km~s$^{-1}$}
\newcommand\alphaco{$\alpha_{\rm CO}$}
\newcommand\co{CO(1-0)}
\newcommand\cotwo{CO(2-1)}
\newcommand\carbon{[CI](1-0)} 

\newcommand\aastex{AAS\TeX}

\received{June 29, 2018}
\revised{June 29, 2018}
\accepted{July 11, 2018}
\submitjournal{ApJ}
\shorttitle{\aastex\ ALMA [CI]$^{3}P_{1}-^{3}P_{0}$ observations of NGC~6240}
\shortauthors{Cicone et al.}

\begin{document}

\title{ALMA [CI]$^{3}P_{1}-^{3}P_{0}$ observations of NGC~6240: a puzzling molecular outflow, and the role of outflows in the global \alphaco~factor of (U)LIRGs}

\correspondingauthor{Claudia Cicone}
\email{claudia.cicone@brera.inaf.it}

\author[0000-0003-0522-6941]{Claudia Cicone}
\altaffiliation{Marie Sk\l{}odowska-Curie fellow}
\affiliation{INAF - Osservatorio Astronomico di Brera, Via Brera 28, 20121 Milano, Italy}

\author{Paola Severgnini} 
\affiliation{INAF - Osservatorio Astronomico di Brera, Via Brera 28, 20121 Milano, Italy}

\author{Padelis P. Papadopoulos} 
\affiliation{Department of Physics, Section of Astrophysics, Astronomy and Mechanics, Aristotle University of Thessaloniki, Thessaloniki, Macedonia, 54124, Greece}
\affiliation{Research Center for Astronomy, Academy of Athens, Soranou Efesiou 4, GR-115 27 Athens, Greece}
\affiliation{School of Physics and Astronomy, Cardiff University, Queen's Buildings, The Parade, Cardiff, CF24 3AA, UK}

\author{Roberto Maiolino} 
\affiliation{Cavendish Laboratory, University of Cambridge, 19 J. J. Thomson Ave., Cambridge CB3 0HE, UK}
\affiliation{Kavli Institute of Cosmology Cambridge, Madingley Road, Cambridge CB3 0HA, UK}

\author{Chiara Feruglio} 
\affiliation{INAF - Osservatorio Astronomico di Trieste, via G.B. Tiepolo 11, 34143 Trieste, Italy}

\author{Ezequiel Treister} 
\affiliation{Instituto de Astrofisica, Facultad de Fisica, Pontificia Universidad Catolica de Chile, Casilla 306, Santiago 22, Chile}

\author{George C. Privon} 
\affiliation{Department of Astronomy, University of Florida, 211 Bryant Space Sciences Center, Gainesville, 32611 FL, USA}

\author{Zhi-yu Zhang} 
\affiliation{European Southern Observatory, Karl-Schwarzschild-Straße 2, 85748 Garching, Germany}
\affiliation{Institute for Astronomy, University of Edinburgh, Royal Observatory, Blackford Hill, Edinburgh EH9 3HJ, UK}

\author{Roberto Della Ceca} 
\affiliation{INAF - Osservatorio Astronomico di Brera, Via Brera 28, 20121 Milano, Italy}

\author{Fabrizio Fiore} 
\affiliation{INAF - Osservatorio Astronomico di Roma, via Frascati 33, 00078 Monteporzio Catone, Italy}

\author{Kevin Schawinski} 
\affiliation{Institute for Particle Physics and Astrophysics, ETH Zurich, Wolfgang-Pauli-Str. 27, CH-8093 Zurich, Switzerland}

\author{Jeff Wagg} 
\affiliation{SKA Organisation, Lower Withington Macclesfield, Cheshire SK11 9DL, UK}


%
%




\begin{abstract}
We present Atacama large millimeter/~submillimeter array (ALMA) and compact array (ACA) [CI]$^{3}P_{1}-^{3}P_{0}$ (\carbon) observations of NGC~6240, which we combine with ALMA \cotwo~and IRAM Plateau de Bure Interferometer \co~data to study the physical properties of the massive molecular (H$_2$) outflow. 
We discover that the receding and approaching sides of the H$_2$ outflow, aligned east-west, exceed 10~kpc in their total extent. 
High resolution ($0.24\arcsec$) \carbon~line images surprisingly reveal that the outflow emission peaks {\it between} the two active galactic nuclei (AGN), rather than on either of the two, and that it dominates the velocity field in this nuclear region.
We combine the \carbon~and \co~data to constrain the CO-to-H$_2$ conversion factor (\alphaco) in the outflow, which is on average $2.1\pm1.2~\rm M_{\odot} (K~km~s^{-1}~pc^2)^{-1}$. We estimate that $60\pm20$~\% of the total H$_2$ gas reservoir of NGC~6240 is entrained in the outflow, for a resulting mass-loss rate of $\dot{M}_{out}=2500\pm1200~M_{\odot}~yr^{-1} \equiv 50\pm30$~SFR. This energetics rules out a solely star formation-driven wind, but the puzzling morphology challenges a classic radiative-mode AGN feedback scenario. For the quiescent gas we compute $\langle\alpha_{\rm CO}\rangle = 3.2\pm1.8~\rm M_{\odot} (K~km~s^{-1}~pc^2)^{-1}$, which is at least twice the value commonly employed for (ultra) luminous infrared galaxies ((U)LIRGs). We observe a tentative trend of increasing $r_{21}\equiv L^{\prime}_{\cotwo}/L^{\prime}_{\co}$ ratios with velocity dispersion and measure $r_{21}>1$ in the outflow, whereas $r_{21}\simeq1$ in the quiescent gas. We propose that molecular outflows are the location of the warmer, strongly unbound phase that partially reduces the opacity of the CO lines in (U)LIRGs, hence
driving down their global \alphaco~and increasing their $r_{21}$ values. 
\end{abstract}




\section{Introduction} \label{sec:intro}

Massive ($M_{mol}>10^8~M_{\odot}$) and extended ($r\gtrsim1$~kpc) outflows of cold and dense molecular (H$_2$) gas have been discovered in a large number of starbursts and active galactic nuclei (AGNs) \citep{Turner85,Nakai+87,Sakamoto+06_NGC3256,Fischer+10,Feruglio+10,Sturm+11,Alatalo+11,Dasyra+Combes12,Veilleux+13,Spoon+13,Combes+13, Morganti+13_IC5063, Feruglio+13a, Cicone+14, Garcia-Burillo+14, Garcia-Burillo+15, Zschaechner+16, Feruglio+17, Carniani+17, Barcos-Munoz+18,
Gowardhan+18, Fluetsch+18}. 
Although so far limited mostly to local (ultra) luminous infrared galaxies ((U)LIRGs), these observations indicate that the mass-loss rates of H$_2$ gas are higher compared to the ionised gas phase participating in the outflows \citep{Carniani+15,Fiore+17}. Therefore, molecular outflows, by displacing and perhaps removing the fuel available for star formation,  can have a strong impact on galaxy evolution. More luminous AGNs host more powerful H$_2$ winds, suggesting a direct link between the two \citep{Cicone+14}. 

The presence of massive amounts of cold and dense H$_2$ gas outflowing at $v\gtrsim1000$~\kms across kpc scales in galaxies is itself puzzling. A significant theoretical effort has gone into reproducing the properties of multiphase outflows in the context of AGN feedback models \citep{Cicone+18}. In one of the AGN radiative-mode scenarios, the outflows result from the interaction of fast highly ionised winds launched from the pc-scales with the kpc-scale interstellar medium (ISM), which occurs through a `blast-wave' mechanism \citep{Silk+Rees98, King10, Zubovas+King12, Faucher-Giguere+12, Costa+14, Gaspari+17, Biernacki+18}. In this picture, because molecular clouds overtaken by a hot and fast wind are quickly shredded \citep{Bruggen+Scannapieco16},  it is more likely that the high-velocity H$_2$ gas forms directly within the outflow, by cooling out of the warmer gas \citep{Zubovas+King14,Costa+15,Nims+15,Thompson+16,Richings+18}. An alternative scenario, not requiring shockwaves, is the direct acceleration of the molecular ISM by radiation pressure on dust  \citep{Thompson+15,Ishibashi+Fabian15, Costa+18}. This mechanism is most efficient in AGNs deeply embedded in a highly IR optically thick medium, such as local (U)LIRGs. 

In order to advance our theoretical understanding of galactic-scale molecular outflows, we need to place more accurate constraints on their energetics. Indeed, most current H$_2$ outflow mass estimates are based on a single molecular gas tracer (CO or OH), implying uncertainties of up to one order of magnitude \citep{Veilleux+17,Cicone+18}. The luminosity of the \co~line, which is optically thick in typical conditions of molecular clouds, can be converted into H$_2$ mass through an \co-to-H$_2$ conversion factor (\alphaco) calibrated using known sources and dependent on the physical state of the gas. 
For the molecular ISM of isolated (or only slightly perturbed) disk galaxies like the Milky Way, the conventional \alphaco~is $4.3~\rm M_{\odot} (K~km~s^{-1}~pc^2)^{-1}$ \citep{Bolatto+13}. Instead, for merger-driven starbursts like most (U)LIRGs, which are characterised by a more turbulent and excited ISM, a lower \alphaco~of $\sim0.6-1.0~\rm M_{\odot} (K~km~s^{-1}~pc^2)^{-1}$ is often adopted \citep{Downes+Solomon98,Yao+03,Israel+15}. 
Such low \alphaco~values have been ascribed to the existence, in the inner regions of these mergers, of a warm and turbulent `envelope' phase of H$_2$ gas, not contained in self-gravitating clouds \citep{Aalto+95}.
However, some recent analyses of the CO spectral line energy distributions (SLEDs) including high-$J$ ($\gtrsim3$) transitions suggest that near-Galactic \alphaco~values are also possible for (U)LIRGs, especially when a significant H$_2$ gas fraction is in dense, gravitationally-bound states \citep{Papadopoulos+12_ApJ}. Dust-based ISM mass measurements also deliver galactic-type \alphaco~factors for (U)LIRGs, although they depend on the underlying assumptions used to calibrate the conversion \citep{Scoville+16}.

Molecular outflows can be significantly fainter than the quiescent ISM, and so multi-transition observations aimed at estimating their \alphaco~are particularly challenging. \cite{Dasyra+16} and more recently \cite{Oosterloo+17}, for the radio-jet driven outflow in IC~5063, derived a low optically-thin \alphaco~of $\sim0.3~\rm M_{\odot} (K~km~s^{-1}~pc^2)^{-1}$, in line with theoretical predictions by \cite{Richings+18}. 
On the other hand, for the starburst-driven M82 outflow, \cite{Leroy+15} calculated\footnote{By using the \cotwo~transition} $\alpha_{CO}^{2-1}=1-2.5~\rm M_{\odot} (K~km~s^{-1}~pc^2)^{-1}$.
The detection of high density gas in the starburst-driven outflow of NGC~253 would also favour an \alphaco~higher than the optically thin value \citep{Walter+17}, and a similar conclusion may be reached for the outflow in Mrk~231, found to entrain a substantial amount of dense H$_2$ gas \citep{Aalto+12,Aalto+15,Cicone+12,Lindberg+16}.

An alternative method for measuring the molecular gas mass, independent of the \alphaco~factor, is through a tracer such as the $^{3}P_{1}-^{3}P_{0}$ transition of neutral atomic carbon (hereafter \carbon). This line, optically thin in most cases, has an easier partition function than molecules and excitation requirements similar to \co\footnote{\co~and \carbon~are similar in critical density but $E_{10}/k_b$=5.5~K for CO and 23~K for [CI]; however, as long as most of the H$_2$ gas has $T_k>15-20$~K, as expected, the $E/k_b$ difference between the two lines makes no real excitation difference in the level population \citep{Papadopoulos+04}.}. More importantly [CI] is expected to be fully coexisting with H$_2$ \citep{Papadopoulos+04, Papadopoulos+Greve04}. Therefore, by combining the information from \co~and \carbon~it is possible to derive an estimate of the \alphaco~value.
Similar to any optically thin species used to trace $\rm H_2$ (e.g. dust, $^{13}$CO), 
converting the \carbon~line flux into a mass measurement is plagued by the unavoidable uncertainty on its abundance. However, in this regard, recent calculations found not only that the {\it average} [C/H$_2$] abundance in molecular clouds is more robust than that of molecules such as CO, but also that [CI] can even trace the H$_2$ gas where CO has been severely depleted by cosmic rays (CRs, \citealt{Bisbas+15,Bisbas+17}).

In this work we use new Atacama large millimeter/~submillimeter array (ALMA) and Atacama compact array (ACA) observations of the \carbon~line in NGC~6240 to constrain the physical properties of its molecular outflow. NGC~6240 is a merging LIRG hosting two AGNs with quasar-like luminosities \citep{Puccetti+16}. The presence of a molecular outflow was suggested by \cite{vanderWerf+93} based on the detection of high-velocity wings of the ro-vibrational H$_2$ $v=1-0$~S(1)~$2.12\mu$m line and by \cite{Iono+07} based on the CO(3-2) kinematics, and it was later confirmed by \cite{Feruglio+13a} using IRAM PdBI \co~observations. 
This is one of the first interferometric \carbon~observations of a local galaxy (see also \citealt{Krips+16}), and - to our knowledge - the first spatially-resolved \carbon~observation of a molecular outflow in a quasar. Probing the capability of \carbon~to image molecular outflows is crucial: besides being an alternative H$_2$ tracer independent of the \alphaco~factor, \carbon~is also sensitive to CO-poor gas, which may be an important component of molecular outflows exposed to strong far-ultraviolet (UV) fields \citep{Wolfire+10} or CR fluxes \citep{Bisbas+15,Bisbas+17,Krips+16,Gonzalez-Alfonso+18}. Moreover, testing \carbon~as a sensitive molecular probe in a local and well-studied galaxy such as NGC~6240 has a great legacy value for studies at $z>2$, where the [CI] lines are very valuable tracers of the bulk of the molecular gas accessible with ALMA \citep{Zhang+16}.

The paper is organised as follows: 
 in $\S$~\ref{sec:obs} we describe the data; in $\S$~\ref{sec:results_outflowmap} we present the 
\co, \cotwo, and \carbon~outflow maps and the \carbon~line moment maps. In $\S$~\ref{sec:results_total}-$\S$~\ref{sec:results_boxes} we identify the outflowing components of the molecular line emission and derive the \alphaco~and $r_{21}$ values separately for the quiescent ISM and the outflow. The outflow energetics is constrained in $\S$~\ref{sec:of_prop}. In $\S$~\ref{sec:alphaco_r21} we study the variations of \alphaco~and $r_{21}$ as a function of $\sigma_v$ and distance of the different spectral components from the nucleus. Our findings are discussed in $\S~\ref{sec:discussion}$ and summarised in $\S~\ref{sec:conclusion}$.
Throughout the paper we adopt a standard $\Lambda$CDM cosmological model with $H_0$ = 67.8 \kms Mpc$^{-1}$, 
$\Omega_{\Lambda}$ = 0.692, $\Omega_{\rm M}$ = 0.308 \citep{Planck2016}. At the distance of NGC~6240 (redshift $z=0.02448$, luminosity distance $D_L = 110.3$~Mpc), the physical scale is 0.509~kpc~arcsec$^{-1}$. Uncertainties correspond to $1\sigma$ statistical errors. The units of \alphaco~[$\rm M_{\odot} (K~km~s^{-1}~pc^2)^{-1}$] are sometimes omitted.

\section{Observations} \label{sec:obs} 

The Band~8 observations of the \carbon~($\nu_{\rm rest}^{\rm [CI]}=492.16065$~GHz) emission line in NGC~6240 
were carried out in May 2016 with the 12~m-diameter antennas of ALMA and in August 2016 with the 7~m-diameter antennas of the Atacama Compact Array (ACA) as part of our Cycle~3 programme 2015.1.00717.S (PI: Cicone). 
The ALMA observations were effectuated in a compact configuration with 40 antennas (minimum and maximum baselines, $b_{min}=15$~m, $b_{max}=640$~m), yielding an angular resolution (AR) of $0.24\arcsec$ and a maximum recoverable scale (MRS) of $2.48\arcsec$. Only one of the two planned 0.7~h-long scheduling blocks was executed, and the total on-source time was 0.12~h. The PWV was 0.65~mm and the average system temperature was $T_{sys}=612$~K.
The ACA observations were performed using nine antennas with $b_{min}=9$~m and $b_{max}=45$~m, resulting in AR=$2.4\arcsec$ and MRS=$14\arcsec$. The total ACA observing time was 3.5~h, of which 0.7~h on source. The average PWV and $T_{sys}$ were 0.7~mm and 550~K, respectively. J1751+0939 and Titan were used for flux calibration, J1924-2914 for bandpass calibration, J1658+0741 and J1651+0129 for phase calibration.

We employed the same spectral setup for the ALMA and the ACA observations. Based on previous \co~observations of NGC~6240 \citep{Feruglio+13a,Feruglio+13b}, and on the knowledge of the concurrence of the \carbon~and \co~emissions, we expected the \carbon~line to be significantly broad (full width at zero intensity, FWZI$>1000$~\kms). Therefore, to recover both the broad \carbon~line and its adjacent continuum, we placed two spectral windows at a distance of 1.8~GHz, overlapping by 75~MHz in their central 1.875~GHz-wide full sensitivity part, yielding a total bandwidth of 3.675~GHz (2293 \kms) centred at $\nu_{obs}^{\rm [CI]}=480.40045$~GHz.

After calibrating separately the ALMA and ACA datasets with the respective scripts delivered to the PI, we fit and subtracted the continuum in the {\it uv} plane. This was done through the \texttt{CASA}\footnote{Common Astronomy Software Applications \citep{McMullin07}.} task \texttt{uvcontsub}, by using a zeroth-order polynomial for the fit, and by estimating the continuum emission in the following line-free frequency ranges: 478.568$<\nu_{obs} [\rm GHz]<$478.988 and 481.517$<\nu_{obs}[\rm GHz]<$482.230. 
The line visibilities were then deconvolved using \texttt{clean} with Briggs weighting and robust parameter equal to 0.5. A spectral binning of $\Delta \nu=9.77$~MHz (6~\kms) was applied, and the cleaning masks were chosen interactively. 
In order to improve the image reconstruction, following the strategy adopted by \cite{Hacar+18}, we used the cleaned ACA data cube corrected for primary beam as a source model to initialise the deconvolution of the ALMA line visibilities (parameter `\texttt{modelimage}' in \texttt{clean}).
The synthesised beams of the resulting ACA and ALMA image data cubes are $4.55\arcsec\times2.98\arcsec$ (PA=$-53.57$ deg) and $0.29\arcsec\times0.24\arcsec$ (PA= 113.18 deg), respectively.  
In addition, a lower resolution ALMA \carbon~line data cube was produced by applying a tapering (outer taper of $1.4\arcsec$), which resulted in a synthesised beam of $1.28\arcsec\times1.02\arcsec$ (PA= 77.49 deg). Primary beam correction was applied to all datasets. 
We checked the accuracy of the ALMA and ACA relative flux scales by comparing the flux on the overlapping spatial scales between the two arrays, and found that they are consistent within the Band~8 calibration uncertainty of 15\%.

As a last step, in order to maximise the uv coverage and the sensitivity to any extended structure possibly filtered out by the ALMA data, we combined the (tapered and non) ALMA image data cubes with the ACA one using the task \texttt{feather}. 
For a detailed explanation of \texttt{feather} we refer the reader to the CASA cookbook\footnote{Available at \texttt{https://casa.nrao.edu}}.
The same steps were used to produce all the interferometric maps shown in this paper, i.e.: (i) \texttt{clean} of ACA visibilities followed by primary beam correction, (ii) \texttt{clean} of ALMA visibilities by using the ACA images as a source model followed by primary beam correction, and (iii) \texttt{feather} of the ACA and ALMA images.
The resulting ACA+ALMA merged images inherit the synthesised beam and cell size  (the latter set equal to $0.01\arcsec$ and $0.2\arcsec$ respectively for the higher and lower resolution data) of the corresponding input ALMA images. 
At the phase tracking centre (central beam), the $1\sigma$ sensitivities to line detection, calculated using the line-free spectral channels, are 1.4~mJy~beam$^{-1}$ and 5~mJy~beam$^{-1}$ per $dv=50$~\kms spectral channel, respectively for the higher and lower resolution \carbon~line data cubes. The sensitivity decreases slightly with distance from the phase center. At a radius of 5$\arcsec$, the \carbon~line sensitivities per $dv=50$~\kms spectral channel are 3.3~mJy~beam$^{-1}$ and 5.5~mJy~beam$^{-1}$.

In this paper we make use of the IRAM PdBI \co~data previously presented by \cite{Feruglio+13a, Feruglio+13b}. The \co~line image data cube used in this analysis has a synthesised beam of $1.42\arcsec\times1.00\arcsec$ (PA= 56.89~deg) and a cell size of 0.2$\arcsec$. The \co~$1\sigma$ line sensitivity per $dv=50$~\kms spectral channel is 0.6~mJy~beam$^{-1}$ at the phase center, and 0.65~mJy~beam$^{-1}$ at a $5\arcsec$ radius. 

Our analysis also includes ALMA Band 6 (programme 2015.1.00370.S, PI: Treister) snapshot (one minute on source) observations of NGC~6240 targeting the \cotwo~transition, which were performed in January 2016 (PWV=1.2~mm) using the compact configuration (AR=$1.2\arcsec$, MRS=$10\arcsec$). These observations were executed in support of the long baseline campaign carried out by Treister et al. (in prep). We calibrated the data using the script for PI, estimated the continuum from the line-free spectral ranges (224.106$<\nu_{obs} [\rm GHz]<$224.279 and 225.780$<\nu_{obs} [\rm GHz]<$225.971) and subtracted it in the {\it uv} plane. We deconvolved the line visibilities using \texttt{clean} with Briggs weighting (robust=0.5) and applied a correction for primary beam. The final \cotwo~cleaned data cube has a synthesised beam of $1.54\arcsec\times0.92\arcsec$ (PA= 60.59~deg) and a cell size of 0.2$\arcsec$. The 1$\sigma$ \cotwo~line sensitivity per $dv=50$~\kms channel is 0.80~mJy~beam$^{-1}$ at the phase centre and $1$~mJy beam$^{-1}$ at a 5$\arcsec$ radius.

In the analysis that follows, the comparison between the \co, \cotwo, and \carbon~line tracers in the molecular outflow of NGC~6240 will be done by using the lower resolution ALMA+ACA \carbon~data cube, which matches in angular resolution ($\sim1.2\arcsec$) the IRAM PdBI \co~and ALMA \cotwo~data. 
Unless specified, quoted errors include the systematic uncertainties on the measured fluxes due to flux calibration, which are 10\% for the IRAM PdBI \co~and ALMA \cotwo~data, and 15\% for the ALMA \carbon~line observations.
We report the presence of some negative artefacts, especially in the cleaned \co~and \cotwo~datacubes. These are due 
to the interferometric nature of the observations, which does not allow to properly recover all the faint 
extended emission in a source with a very bright central peak emission such as NGC~6240. However, the 
negative features lie mostly outside the region probed by our analysis and they are not expected to significantly affect our flux recovery, since the total \co~and \cotwo~line fluxes are consistent with previous single-dish measurements \citep{Costagliola+11,Papadopoulos+12_MNRAS}. Our total \carbon~line flux is higher than that recovered by \cite{Papadopoulos+Greve04} by using the James Clerk Maxwell telescope (JCMT, FWHM$_{beam}=10\arcsec$), but lower by $34\pm15$\% than the flux measured by the {\it Herschel} space observatory \citep{Papadopoulos+14}. This indicates that some faint extended emission has been resolved out and/or that there is additional \carbon~line emission outside the field of view of our observations. 

\section{Data analysis and results}

\subsection{Morphology of the extended molecular outflow and its launch region}\label{sec:results_outflowmap} 

\begin{figure*}[ptb] 
\centering
\includegraphics[width=.5\textwidth,angle=90]{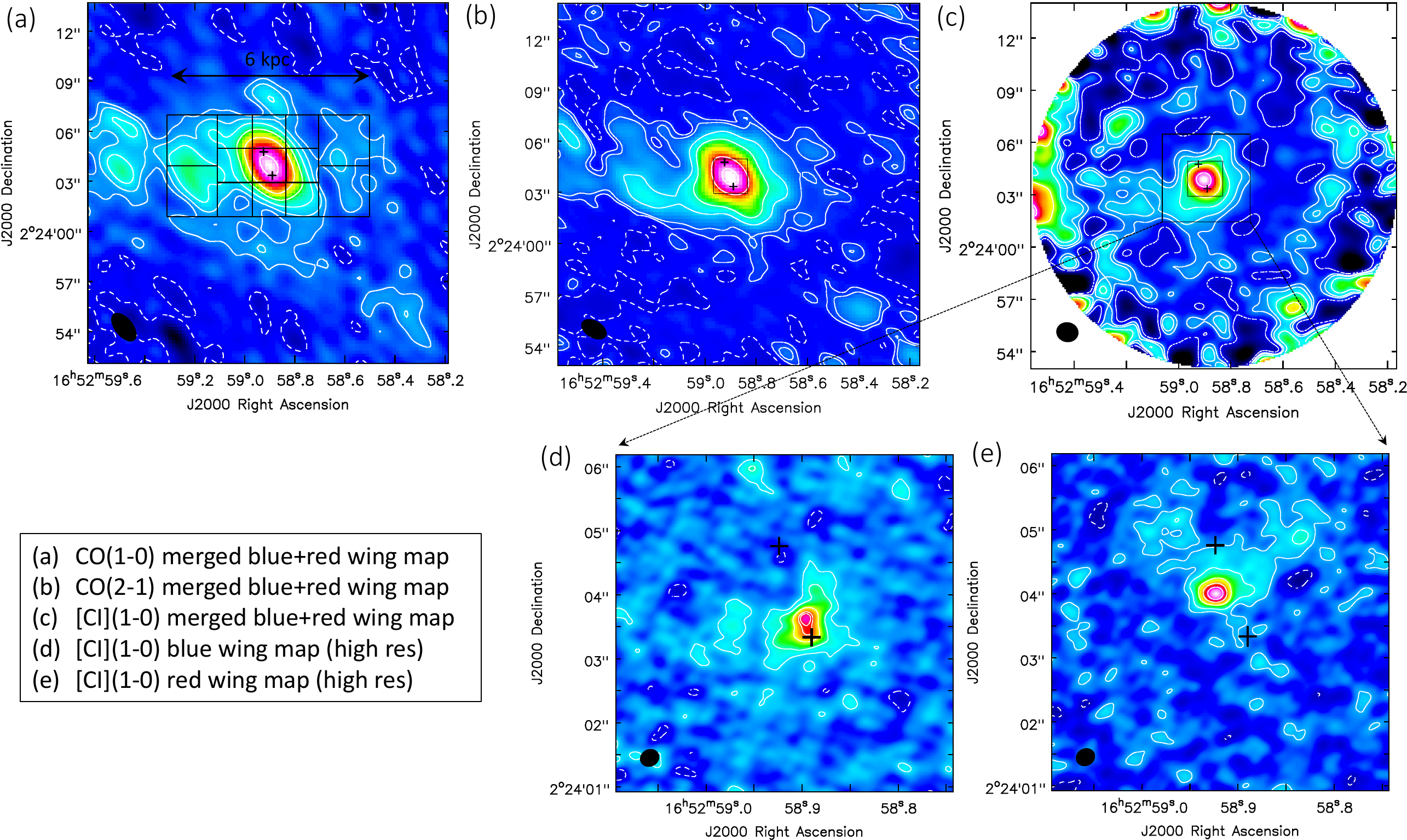}
\caption{The extended NGC~6240 outflow observed using different molecular gas tracers. The outflow emission, integrated within $v\in(-650, -200)$~\kms (blue wing) and $v\in(250, 800)$~\kms (red wing) and combined together, is shown in the maps (a), (b), and (c) respectively for the \co, \cotwo, and \carbon~transitions. The three maps have matched spatial resolution ($\sim1.2\arcsec$, details in $\S$~\ref{sec:obs}). 
Contours correspond to: ($-3\sigma$, $3\sigma$, $6\sigma$, $12\sigma$, $24\sigma$, $48\sigma$, $150\sigma$) with $1\sigma$= 0.14 mJy~beam$^{-1}$ in panel (a); ($-3\sigma$, $3\sigma$, $6\sigma$, $24\sigma$, $48\sigma$, $200\sigma$, $400\sigma$) with $1\sigma$= 0.23~mJy~beam$^{-1}$ in panel (b); 
($-3\sigma$, $3\sigma$, $6\sigma$, $12\sigma$, $24\sigma$, $48\sigma$) with $1\sigma$= 1.23 mJy~beam$^{-1}$ in panel (c). 
Panels (d) and (e) show the maps of the \carbon~blue and red wings at the original spatial resolution of the ALMA Band~8 data ($0.24\arcsec$, details in $\S$~\ref{sec:obs}). Contours correspond to ($-3\sigma$, $3\sigma$, $6\sigma$, $12\sigma$, $18\sigma$, $20\sigma$) with $1\sigma$= 1.1~mJy~beam$^{-1}$ in panel (d) and 
$1\sigma$= 1~mJy~beam$^{-1}$ in panel (e). 
The black crosses indicate the VLBI positions of the AGNs from \cite{Hagiwara+11}. The synthesised beams are shown at the bottom-left of each map. 
The grid encompassing the central $12\arcsec\times 6\arcsec$ region and employed in the spectral analyses presented in $\S$~\ref{sec:results_total} and $\S$~\ref{sec:results_boxes} is drawn in panel (a). }
\label{fig:outflow_maps}
\end{figure*}

\begin{figure}[ptb]  
\centering
\includegraphics[width=.8\columnwidth,angle=90]{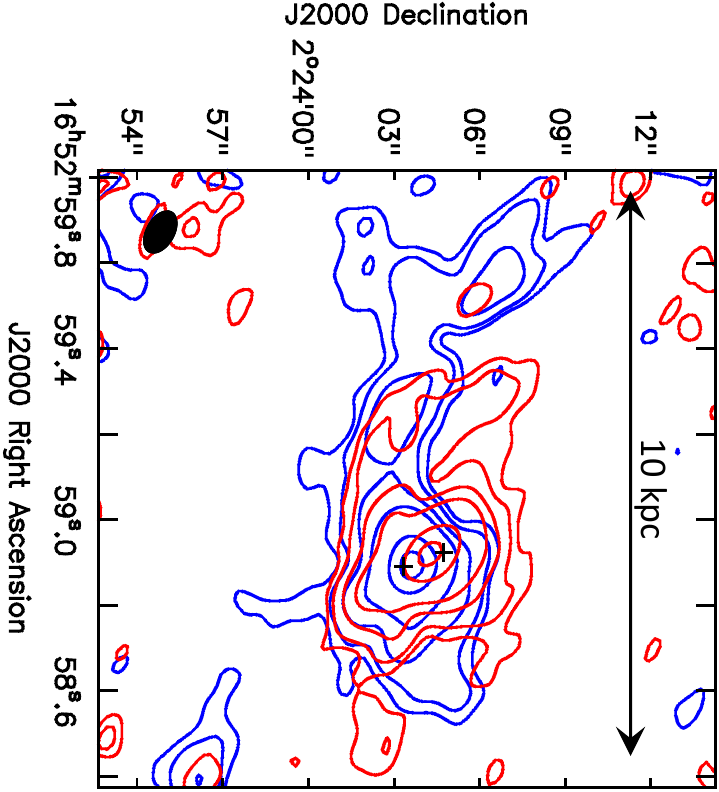}
\caption{Comparison between the \cotwo~blue and red wing emissions in NGC~6240. For visualisation purposes, only positive contours starting from $5\sigma$ are shown, with $1\sigma$= 0.33 mJy~beam$^{-1}$ for the blue wing (blue contours) and $1\sigma$= 0.3 mJy~beam$^{-1}$ for the red wing (red contours). The corresponding interferometric maps including negative contours are displayed in Appendix~\ref{sec:additional_COmaps} (Figure~\ref{fig:COwings}). } 
\label{fig:CO21wings_overlap}
\end{figure}

\begin{figure*}[ptb]  
\centering
\includegraphics[width=.85\textwidth]{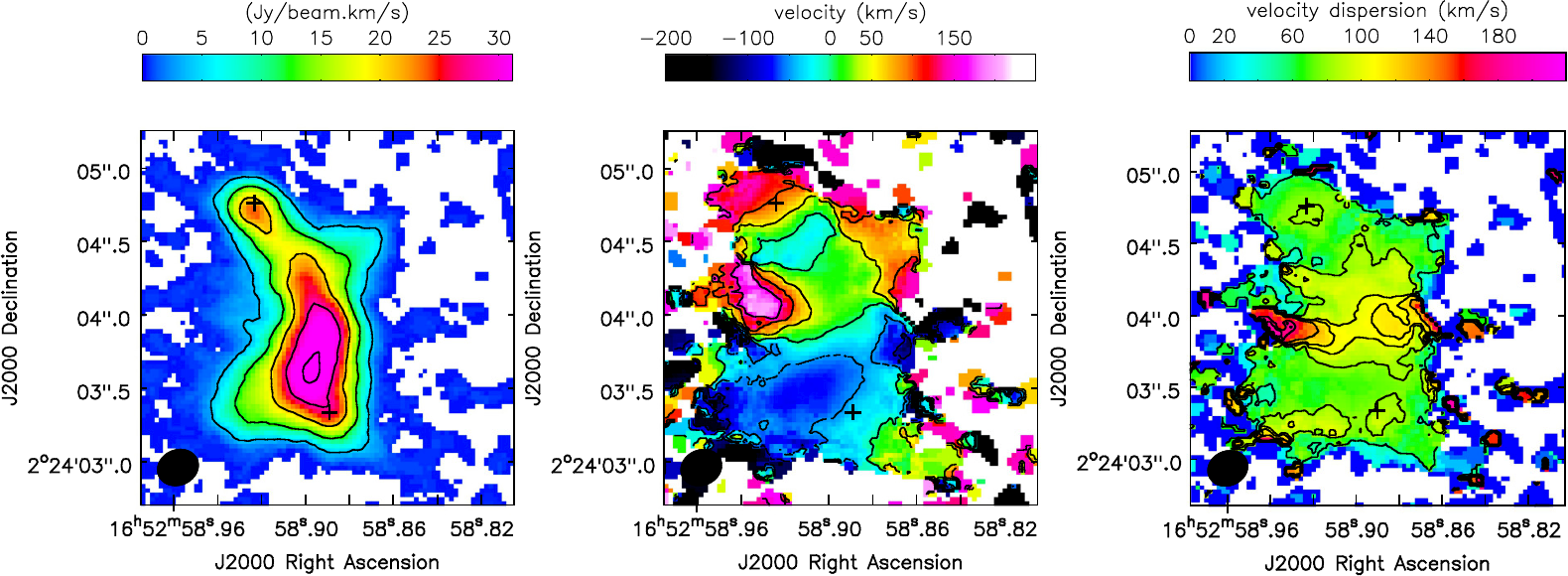}
\caption{Intensity-weighted moment maps of the \carbon~line emission in the merger nucleus. The maps were computed from the higher resolution ALMA+ACA merged data cube (see $\S$~\ref{sec:obs}) by using the task \texttt{immoments} and by selecting the spectral range $v\in(-200, 250)$~\kms. Contours correspond to: [-100, -50, 0, 50, 100, 150]~\kms (moment 1, {\it central panel}) and [50, 80, 100, 110, 130, 170, 180]~\kms (moment 2, {\it right panel}).} 
\label{fig:CI10_moments}
\end{figure*}

With the aim of investigating the extent and morphology of the outflow, we produced interferometric maps of the \co, \cotwo, and \carbon~high-velocity emissions. The maps, shown in Fig.~\ref{fig:outflow_maps}(a,b,c), were generated by merging together and imaging the {\it uv} visibilities corresponding to the blue- and red-shifted wings of the molecular lines, integrated respectively within $v\in(-650, -200)$~\kms and $v\in(250, 800)$~\kms. 
These are the velocity ranges that, following from the identification of the outflow components performed in $\S$~\ref{sec:results_boxes} (and further discussed in $\S$~\ref{sec:disc_caveats} and Appendix~\ref{sec:boxspectra_appendix}), are completely dominated by the emission from outflowing gas. 
The data displayed in panels (a,b,c) have a matched spatial resolution equal to $\sim1.2\arcsec$ (details in $\S$~\ref{sec:obs}). Panels (d,e) of  Fig.~\ref{fig:outflow_maps} show the maps of the blue and red \carbon~line wings at the native spatial resolution of the ALMA Band 8 observations ($0.24\arcsec$, $\S$~\ref{sec:obs}).

The extended ($>5$~kpc) components of the outflow are best seen in Fig.\ref{fig:outflow_maps}(a,b,c), whereas panels (d,e) provide a zoomed view of the molecular wind in the inner 1-2 kpc. The bulk of the outflow extends eastward of the two AGNs, as already pointed out in the previous analysis of the \co~data done by \cite{Feruglio+13a}. In addition, we identify for the first time a western extension of the molecular outflow, roughly aligned along the same east-west axis as the eastern component. At the sensitivity allowed by our data, we detect at a S/N$>5$ CO emission features associated with the outflow up to a maximum distance of $13.3\arcsec$ ($6.8$~kpc) and $7.2\arcsec$ ($3.7$~kpc) from the nucleus, respectively in the east and west directions (Fig~\ref{fig:outflow_maps}a,b,c). The \carbon~map in Fig~\ref{fig:outflow_maps}(c) shows extended structures similar to CO, although the limited field of view of the Band~8 observations does not allow us to probe emission beyond a radius of $\sim7.5\arcsec$.

\cite{Feruglio+13a} hinted at the possibility that the redshifted CO detected in NGC~6240 could be involved in the feedback process, but did not explicitly ascribe it to the outflow because of the smaller spatial extent of the red wing with respect to the blue wing. In Figure~\ref{fig:CO21wings_overlap} we directly compare the blue and red line wings using the ALMA \cotwo~data. Based on their close spatial correspondence, whereby the red wing overlaps with the blue one across more than $7$~kpc along the east-west direction, we conclude that both the red-shifted and blue-shifted velocity components trace the same massive molecular outflow. It follows that the eastern and western sides of the outflow are detected in both their approaching and receding components. At east, the blue-shifted emission is brighter than the redshifted one and dominant beyond a $3.5$~kpc radius. 

The ALMA \carbon~data can be used to identify with high precision the location of the inner portion of the molecular outflow. Figures~\ref{fig:outflow_maps}(d,e) clearly show that the red wing peaks in the midpoint between the two AGNs, and that the blue wing has a maximum of intensity closer to the southern AGN, as already noted by \cite{Feruglio+13a}. However, these data reveal for the first time that {\it neither the blue nor the red-shifted high velocity \carbon~emissions peak exactly at the AGN positions}. The blue wing has a maximum of intensity at RA(J2000) = 16:52:58.8946$\pm0.0011s$, Dec(J2000)=$+02.24.03.52\pm0.02\arcsec$, offset by $0.18\arcsec\pm0.02\arcsec$ to the north-east with respect to the southern AGN. The red wing instead peaks at RA(J2000) = $16:52:58.9224\pm0.0007s$, Dec(J2000)=$+02.24.04.0158\pm0.007\arcsec$, i.e. at an approximately equal distance of $0.8\arcsec$ from the two AGNs. The \carbon~red and blue wing peaks are separated by $0.65\arcsec\pm0.02\arcsec$. This separation is consistent with the distance between the \cotwo~peaks reported by \cite{Tacconi+99}, although in that work they were interpreted as the signature of a rotating molecular gas disk. 

The presence of such nuclear rotating H$_2$ structure has been largely debated in the literature, especially due to the very high CO velocity dispersion in this region ($\sigma>300$~\kms), and to the mismatch between the dynamics of H$_2$ gas and stars \citep{Gerssen+04,Engel+10}.  
Following \cite{Tacconi+99} and \cite{Bryant+Scoville99}, if a rotating disk is present, its signature should appear at {\it lower} projected velocities than those imaged in Fig~\ref{fig:outflow_maps}(d,e). In Figure~\ref{fig:CI10_moments} we show the high resolution intensity-weighted moment maps of the \carbon~line emission within $-200<v [\rm km~s^{-1}]<250$.
The velocity field does not exhibit the characteristic butterfly pattern of a rotating disk, but it presents a highly asymmetric gradient whereby blue-shifted velocities dominate the southern emission, whereas near-systemic and redshifted velocities characterise the northern emission. These features in the velocity field are correlated in both velocity and position with the high velocity wings 
(Figure~\ref{fig:outflow_maps}(d,e)). Furthermore, the right panel of Fig.~\ref{fig:CI10_moments} shows that the velocity dispersion is uniform ($50\lesssim\sigma_v [\rm km~s^{-1}]\lesssim80$) throughout the entire source and enhanced ($\sigma_v\geq100$~\kms) in a central hourglass-shaped structure extending east-west, which is the same direction of expansion of the larger-scale outflow. This structure has a high-$\sigma_v$ peak with $\sigma_v\geq150$~\kms to the east and another possible peak to the west with $\sigma_v\geq130$~\kms. Such high-$\sigma_v$ points coincide with the blue and red-shifted velocity peaks detected in the moment~1 map. Therefore, based on Fig.~\ref{fig:CI10_moments}, we conclude that {\it the molecular gas emission between the two AGNs is dominated by a nuclear outflow expanding east-west and connected to the larger-scale outflow shown in Fig.\ref{fig:outflow_maps}}. The regions of enhanced turbulence may represent the places where the outflow opening angle widens up - hence increasing the line-of-sight velocity dispersion and velocity of the molecular gas. 

\subsection{The \alphaco~values estimated from the integrated spectra: a reference for unresolved studies}\label{sec:results_total}  

\begin{figure}[tbp]
\centering
\includegraphics[clip=true,trim=2cm 3.5cm 1.2cm 3.5cm, width=.49\textwidth,angle=180]{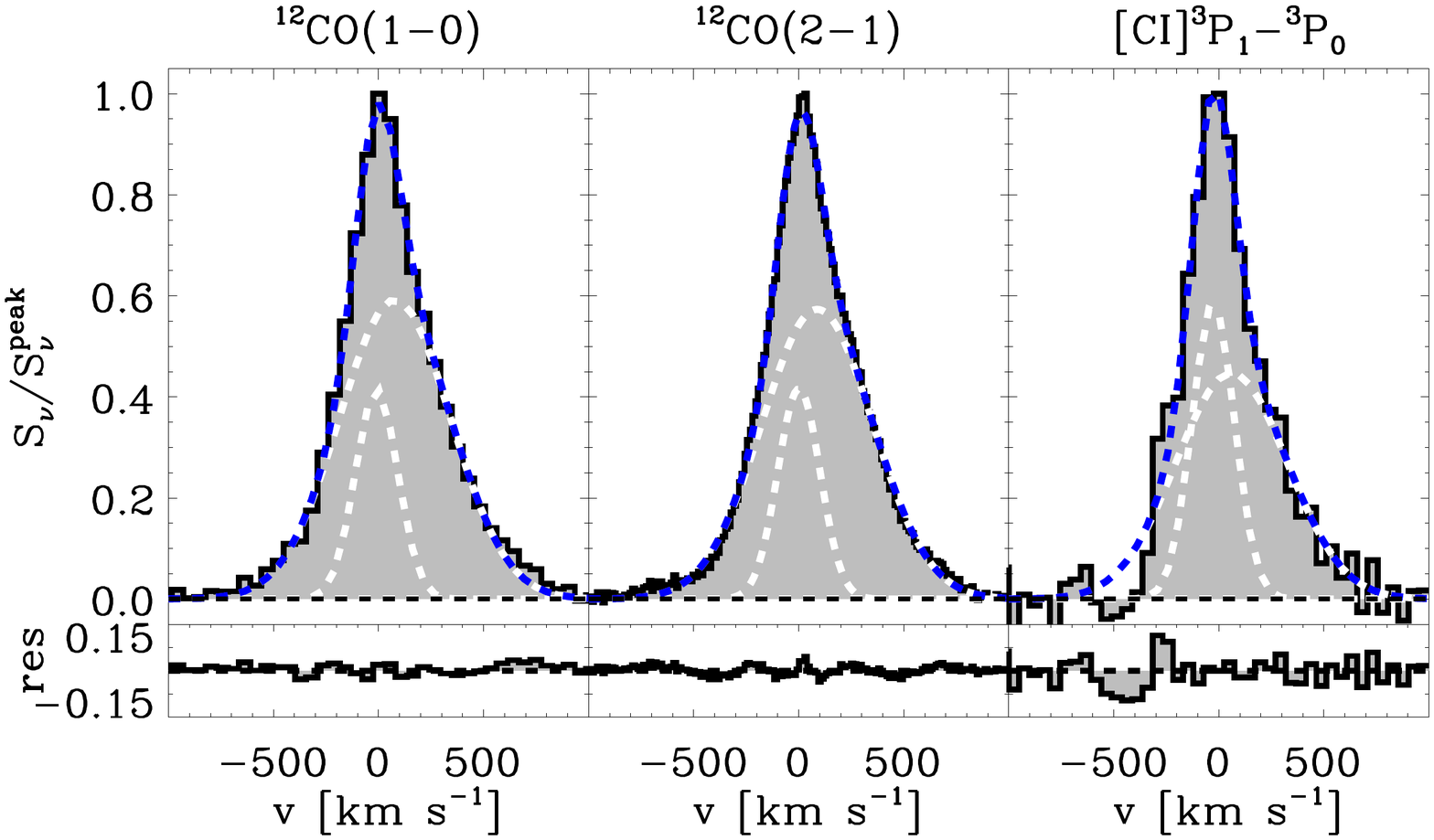}
\caption{Total \co, \cotwo, and \carbon~spectra extracted from a $12\arcsec \times 6\arcsec$-size rectangular region
centred at RA=16:52:58.900, Dec=02.24.03.950 and displayed in Fig.~\ref{fig:outflow_maps}(a). The rms values per spectral channel are: 3.2~mJy ($\delta v=53$~\kms), 17~mJy ($\delta v=13$~\kms) and 78~mJy ($\delta v=49$~\kms), respectively for \co, \cotwo, and \carbon. The spectra were simultaneously fitted using two Gaussian functions (white dashed curves) tied to have the same velocity and width in all three transitions. The best fit results are reported in Table~\ref{table:fit_total}.
The source-averaged $r_{21}$ and \alphaco~calculated from this fit are listed in Table~\ref{table:r21_alphaco_values}.}
\label{fig:tot_region_fit}
\end{figure}

\setcounter{table}{0}
\begin{table}[tbp]
\centering
\scriptsize
\caption{Results of the simultaneous fit to the total spectra$^{*}$}
\label{table:fit_total}
\begin{tabular}{lccc}
\hline
\hline
						& 	\co 	&  \cotwo & \carbon  \\	
\hline
 & \multicolumn{3}{c}{Narrow component} \\
\hline
$v^{\dag}$ [\kms]			&	$-9.1\pm1.0$  & $-9.1\pm1.0$ & $-9.1\pm1.0$ 		       \\
$\sigma_v$ [\kms]	        			&	$101.0\pm1.0$ & $101.0\pm1.0$ & $101.0\pm1.0$		       		\\
$S_{peak}$ [mJy]			& 	$211\pm4$ & $1058\pm12$ & $1360\pm70$		\\
$\int S_{v} dv$ [Jy \kms]		&	$53.4\pm1.0$	& $268\pm 4$	& $344\pm 18$	\\
$L^{\prime}$ [10$^9$ K \kms pc$^2$] &	$1.55\pm 0.03$	& $1.94\pm 0.03$	& $0.55\pm 0.03$  \\
\hline
 & \multicolumn{3}{c}{Broad component}  \\
\hline
$v^{\dag}$ [\kms]			&	$78.0\pm1.2$ & $78.0\pm1.2$ & $78.0\pm1.2$		       \\
$\sigma_v$ [\kms]	        			&	$264.4\pm1.0$ & $264.4\pm1.0$ & $264.4\pm1.0$		       		\\
$S_{peak}$ [mJy]			&	$301\pm3$ 	& $1458\pm12$ 	& 	$1040\pm40$		\\
$\int S_{v} dv$ [Jy \kms]		& 	$199\pm 2$ 	& $966\pm 9$		&   $690\pm 30$	\\
$L^{\prime}$ [10$^9$ K \kms pc$^2$]	& $5.78\pm 0.06$ & $7.01\pm 0.06$ &   $1.09\pm 0.05$  \\
\hline
 & \multicolumn{3}{c}{Total line}  \\
 \hline
$\int S_{v} dv$ [Jy \kms] 		&	$253\pm2$ & $1234\pm10$ & 	$1030\pm30$		\\
$L^{\prime}$ [10$^9$ K \kms pc$^2$]  &	$7.33\pm 0.07$	& $8.96\pm 0.07$	& $1.64 \pm 0.05$	\\
\hline
\end{tabular}

\begin{flushleft}
\small
$^{*}$ The errors quoted in this table are purely statistical and do not include the absolute flux calibration uncertainty. \\
$^{\dag}$ We employ the optical Doppler definition. The fit allows for a global velocity shift of \cotwo~and \carbon~with respect to \co~to take into account the different spectral binning. The best-fit returns: $v_{\rm CO(2-1)}-v_{\rm CO(1-0)} = 10.2 \pm 0.9$~\kms and $v_{\rm [CI](1-0)}-v_{\rm CO(1-0)} = -18 \pm 4$~\kms. \\
\end{flushleft}
\end{table}

Figure~\ref{fig:tot_region_fit} shows the \co, \cotwo, and \carbon~spectra extracted from the $12\arcsec \times 6\arcsec$-size rectangular aperture reported in Fig~\ref{fig:outflow_maps}(a), encompassing both the nucleus and the extended molecular outflow of NGC~6240. The analysis of these integrated spectra, described below, is aimed at deriving a source-averaged \alphaco~for the quiescent and outflowing molecular ISM in NGC~6240. Such analysis is included here because it can be useful as a reference for unresolved observations, for example high redshift analogues of this merger. We stress however that the quality of our data, the proximity of the source, and its large spatial extent allow us to perform a much more detailed, spatially-resolved analysis. The latter will be presented in $\S$~\ref{sec:results_boxes} and delivers the most reliable \alphaco~values for the outflow and the quiescent gas.

The spectra in Fig.~\ref{fig:tot_region_fit} were fitted simultaneously using two Gaussians to account for the narrow core and broad wings of the emission lines, by constraining the central velocity ($v$) and velocity dispersion ($\sigma_v$) of each Gaussian to be equal in the three transitions. Table~\ref{table:fit_total} reports the best-fit results and the corresponding line luminosities calculated from the integrated fluxes following \cite{Solomon+97}. 
The \carbon~line luminosities listed in Table~\ref{table:fit_total} are employed to measure the molecular gas mass ($M_{mol}$, including the contribution from Helium) associated to the narrow and broad line components. The expression for local thermodynamic equilibrium (LTE, i.e. uniform $T_{ex}$) and  
optically thin emission ($\tau_{\rm \carbon} \ll 1$), assuming a negligible background (CMB temperature, $T_{\rm CMB}\ll T_{ex}$) and the Rayleigh-Jeans approximation ($h\nu_{\rm \carbon}\ll kT_{ex}$) is:
\begin{multline}\label{eq:molmass}
M_{mol} [M_{\odot}] = (4.31\mathbin{\cdot}10^{-5}) \mathbin{\cdot} X_{\rm CI}^{-1} \mathbin{\cdot} \bigl(1\mathbin{+} \\
3e^{-23.6/T_{ex}[\rm K]}\mathbin{+} 5e^{-62.5/T_{ex}[\rm K]}\bigr) \mathbin{\cdot} \\
e^{23.6/T_{ex}[\rm K]} \mathbin{\cdot} L^{\prime}_{\rm \carbon} [\rm K~km~s^{-1}~pc^2],
\end{multline}
where $X_{\rm CI}$ is the $\rm [CI/H_2]$ abundance ratio and $T_{ex}$ is the excitation temperature of the gas (see detailed explanations by \cite{Papadopoulos+04} and \cite{Mangum+15}). 
We adopt $T_{ex} = 30$~K and $X_{\rm CI} = (3.0\pm1.5) \times10^{-5}$, which are appropriate for (U)LIRGs \citep{Weiss+03,Weiss+05,Papadopoulos+04,Walter+11,Jiao+17}.
These assumptions will be further discussed in $\S$~\ref{sec:disc_caveats_alphaco}.

By defining a \carbon-to-H$_2$ conversion factor ($\alpha_{\rm [CI]}$) in analogy with the commonly employed \alphaco~factor \citep{Bolatto+13}, Eq~\ref{eq:molmass} resolves into:
\begin{multline}\label{eq:alpha_CI}
M_{mol} [\rm M_{\odot}] \equiv \alpha_{\rm [CI]}~L^{\prime}_{\rm \carbon} [\rm K~km~s^{-1}~pc^2] \\
{\rm with~} 
\alpha_{\rm [CI]} = 9.43~[\rm M_{\odot}~ (K~km~s^{-1}~pc^2)^{-1}].
\end{multline}
Using the values in Table~\ref{table:fit_total} and Eq~\ref{eq:alpha_CI}, we obtain a total molecular gas mass of $M_{mol}=(1.5\pm0.8)\cdot10^{10}~M_{\odot}$, of which $(5\pm3)\cdot10^{9}~M_{\odot}$ is in the narrow component, and $(10\pm5)\cdot10^{9}~M_{\odot}$ in the broad wings.
We then use these $M_{mol}$ values to estimate \alphaco:
\begin{equation}\label{eq:alpha_co}
\alpha_{\rm CO} = M_{mol} [\rm M_{\odot}]~\bigl(L^{\prime}_{\rm \co} [\rm K~km~s^{-1}pc^2]\bigr)^{-1}.
\end{equation}
The results are reported in the first three rows of Table~\ref{table:r21_alphaco_values} for the the total, narrow, and broad emissions in NGC~6240. 
The so-derived \alphaco~factors differ between the narrow and broad line components, being a factor of $1.8\pm0.5$ lower in the latter\footnote{In estimating the error on this ratio we have ignored the systematic uncertainty on $X_{\rm CI}$, assuming it affects both \alphaco~measurements in the same way.}. In the narrow component, the \alphaco~is significantly higher than the typical (U)LIRG value. 
As mentioned in \S~$\ref{sec:intro}$, higher \alphaco~values become possible in (U)LIRGs if a significant fraction of the mass is `hidden' in dense and bound H$_2$ clouds. This is probably the case of NGC~6240, in which a large study using CO SLEDs from $J=1-0$ up to $J=13-12$ from the Herschel space observatory as well as multi-$J$ HCN, CS, and HCO$^{+}$ line data from ground-based observatories, finds \alphaco$\sim 2-4~\rm M_{\odot}~ (K~km~s^{-1}~pc^2)^{-1}$ \citep{Papadopoulos+14}, consistent with our estimates.

Table~\ref{table:r21_alphaco_values} lists also the \cotwo/\co~luminosity ratios, defined as
\begin{equation}\label{eq:r21}
r_{21}\equiv L^{\prime}_{\rm CO(2-1)}/L^{\prime}_{\rm CO(1-0)}.
\end{equation}
We find $r_{21}$ consistently $\sim1.2$ - hence higher than unity (at the $1.5\sigma$ level) - for both the narrow and broad Gaussian components. 
Since our total \co~and \cotwo~fluxes are consistent with previous measurements \citep{Papadopoulos+12_MNRAS, Costagliola+11, Saito+18}, we exclude that 
spatial filtering due to an incomplete {\it uv} coverage is significantly affecting the $r_{21}$ values.
As pointed out by \cite{Papadopoulos+12_MNRAS}, galaxy-averaged $r_{21}>1$ values are not uncommon in (U)LIRGs and are indicative of extreme gas conditions. In this analysis of the integrated spectra we derived $r_{21}>1$ in both the broad and narrow components. However, as we will show in $\S$~\ref{sec:results_boxes} and $\S$~\ref{sec:alphaco_r21}, the spatially-resolved analysis will reveal that $r_{21}\gtrsim1$ values are typical of the outflowing gas and in general of higher-$\sigma_v$ components, while the `quiescent' ISM has $r_{21}\sim1$. 

\subsection{Spatially-resolved analysis: the average \alphaco~of the quiescent and outflowing gas}\label{sec:results_boxes}

\setcounter{table}{1}
\begin{table}[tbp]
\centering
\scriptsize
\caption{\alphaco~and $r_{21}$ values$^*$} 
\label{table:r21_alphaco_values}
\begin{tabular}{lcc}
\hline
\hline
						&  \alphaco 							& $r_{21}$ \\	
						& $[\rm M_{\odot} (K~km~s^{-1}~pc^2)^{-1}]$    &   \\
\hline
Total$^\dag$ 				& 	$2.1\pm 1.1$ & 	$1.22\pm 0.14$   \\
Total$^\dag$ narrow comp 	& 	$3.3\pm 1.8$ &  	$1.25\pm 0.18$ \\
Total$^\dag$ broad comp 		&   	$1.8 \pm 0.9$ &      $1.21\pm 0.17$  \\
\hline
Mean$^\ddag$ global 		&  	$2.5 \pm 1.4$  &   $1.17\pm 0.19$  	\\
Mean$^\ddag$ systemic comp 	&   	$3.2\pm 1.8$ &   $1.0\pm 0.2$  \\
Mean$^\ddag$ outflow comp	&    	$2.1 \pm 1.2$ &   $1.4\pm 0.3$   	   	\\
\hline
\end{tabular}

\begin{flushleft}
\small
$^*$ Quoted errors are dominated by systematic uncertainties (e.g. absolute flux calibration errors, error on $X_{\rm CI}$).\\
$^\dag$ Calculated from the simultaneous fit to the total \co, \cotwo, and \carbon~spectra shown in Fig.~\ref{fig:tot_region_fit}, whose results are reported in Table~\ref{table:fit_total} (details in $\S$~\ref{sec:results_total}).\\
$^\ddag$ Mean values calculated from the simultaneous fit to the \co, \cotwo, and \carbon~spectra extracted from the grid of 13 boxes shown in Fig~\ref{fig:outflow_maps}(a), as explained in $\S$~\ref{sec:results_boxes}. The corresponding spectral fits are shown in Appendix~\ref{sec:boxspectra_appendix} (Figs~\ref{fig:boxes_fit_co10}, \ref{fig:boxes_fit_co21}, \ref{fig:boxes_fit_ci10}).  
\end{flushleft}
\end{table}

\begin{figure}[tb]
\centering
\includegraphics[clip=true,trim=9cm 4.5cm 1cm 3.5cm,width=.8\columnwidth,angle=180]{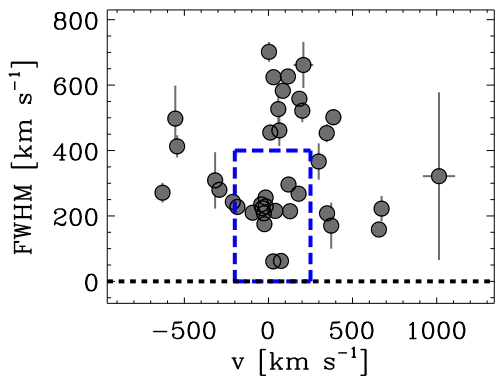}\\
\caption{FWHM as a function of central velocity of all Gaussian components employed in the simultaneous fitting of the \co, \cotwo, and \carbon~box spectra. The blue dashed rectangle constrains the region of the parameter space that we ascribe to the `systemic' components. 
}
\label{fig:v_sigma_boxes}
\end{figure}

The previous analysis ($\S$~\ref{sec:results_total}) was based on the spectral fit shown in Fig.~\ref{fig:tot_region_fit}, where we decomposed the total molecular line emission into a narrow and a broad Gaussian. In first approximation, these two spectral components can be respectively identified with the quiescent and outflowing molecular gas reservoirs of NGC~6240. However, the superb S/N and spatial resolution of our data allow us to take this analysis one step further and refine the definition of quiescent and outflowing components. This is done by including the spatially-resolved information provided by the interferometric data, as described below. 

We divide the central $12\arcsec\times6\arcsec$ region employed in the previous analysis into a grid of 13 squared boxes and use them as apertures to extract the corresponding \co, \cotwo, and \carbon~spectra. As shown in Fig.~\ref{fig:outflow_maps}(a), the central nine boxes have a size of $2\arcsec \times 2\arcsec$, while the external four boxes have a size of $3\arcsec \times 3\arcsec$. The box spectra are presented in Appendix~\ref{sec:boxspectra_appendix} (Figures~\ref{fig:boxes_fit_co10}, \ref{fig:boxes_fit_co21}, and~\ref{fig:boxes_fit_ci10}).

For each box, the \co, \cotwo, and \carbon~spectra are fitted simultaneously with a combination of Gaussian functions tied to have the same line centres and widths for all three transitions. In the fitting procedure, we minimise the number of spectral components required to reproduce the line profiles, up to a maximum of four Gaussians per box. The Gaussian functions employed by the simultaneous fit span a wide range in FWHM and velocity, shown in Figure~\ref{fig:v_sigma_boxes}. The next step is to classify each of these components as `systemic' or `outflow'. In many local (U)LIRGs molecular outflows can be traced through components whose kinematical and spatial features deviate from a rotating molecular structure \citep{Cicone+14,Garcia-Burillo+14}. However, in this source we do not detect any clear velocity gradient that may indicate the presence of a rotating molecular gas disk (Figure~\ref{fig:CI10_moments}).  Therefore, we adopt a different method and identify as `quiescent' the gas probed by the spectral narrow line components that are detected throughout the entire source extent (Figures~\ref{fig:boxes_fit_co10}, \ref{fig:boxes_fit_co21}, and~\ref{fig:boxes_fit_ci10}). Our simultaneous fit to the \co, \cotwo, and \carbon~spectra returns for these narrow components typical FWHM and central velocities in the ranges: FWHM$<400$~\kms and $-200<v[km~s^{-1}]<250$, consistent with what found by \cite{Feruglio+13b}\footnote{\cite{Feruglio+13b} analysed the \co~spectra extracted from different positions within NGC~6240 and found maximum velocity shift and FWHM of the narrow Gaussians of $|v^{max}_{sys}| = 82\pm4$~\kms and FWHM$^{max}_{sys} = 380\pm150$~\kms.}. Based on these results, we assume that all components with $-200<v[km~s^{-1}]<250$ {\it and} FWHM$<400$~\kms trace quiescent gas that is not involved in the outflow. These constraints correspond to the region of the FWHM-$v$ parameter space delimited by the blue-dashed lines in Fig.~\ref{fig:v_sigma_boxes}. 
All components outside this rectangular area are classified as `outflow'. These assumptions are discussed in detail and validated in Appendix~\ref{sec:boxspectra_appendix}, whereas more general considerations about our outflow
identification method are reported in $\S$~\ref{sec:disc_caveats_outflowid}.

Using the results of the simultaneous fit, we measure, for each box and for each of the \co, \cotwo, and \carbon~transitions, the velocity-integrated fluxes apportioned in the `systemic' and `outflow' components. These are computed by summing the fluxes from the respectively classified Gaussian functions fitted to the molecular line profiles.  
For example, in the case of the central box (labelled as `\texttt{C1}' in Figs.~\ref{fig:boxes_fit_co10}, \ref{fig:boxes_fit_co21}, and~\ref{fig:boxes_fit_ci10}), the simultaneous fit employs three Gaussians: a narrow one  
classified as `systemic', and two additional ones classified as `outflow'. The flux of the first Gaussian corresponds to the flux of the `systemic' component for this box, whereas the total `outflow' component flux is given by the sum of the fluxes of the other two Gaussians (the errors are added in quadrature).  

The velocity-integrated fluxes (total, systemic, outflow) are then converted into line luminosities and, from these, the corresponding \alphaco~and $r_{21}$ can be derived by following the same steps as in $\S$~\ref{sec:results_total} (Eq.~\ref{eq:alpha_CI}-\ref{eq:r21}).
Table~\ref{table:r21_alphaco_values} (bottom three rows) lists the resulting mean values of \alphaco~and $r_{21}$ obtained from the analysis of all 13 boxes. In computing the mean, we only include the components detected at a $S/N\geq3$ in each of the transitions used to calculate \alphaco~or $r_{21}$, i.e. \co~and \carbon~for the former, and \co~and \cotwo~for the latter. 
The new \alphaco~values derived for the systemic and outflowing components, respectively equal to $3.2\pm1.8$ and $2.1\pm1.2$, are perfectly consistent with the previous analysis based on the integrated spectra.
Instead, this new analysis delivers different $\langle r_{21}\rangle$ values for the systemic ($1.0\pm0.2$) and outflowing component ($1.4\pm0.3$), although still consistent if considering the associated uncertainties (dominated by the flux calibration errors). 

By using the \co~line data and summing the contribution from all boxes, including both the systemic and the outflowing components, we derive a total molecular gas mass of ${M}_{mol}^{tot} = (2.1\pm 0.5)\times10^{10}~\rm M_{\odot}$. To compute the ${M}_{mol}$ within each box we adopt, when available, the `global' \alphaco~factor estimated for that same box, otherwise we use the mean value of \alphaco=$2.5\pm1.4$\footnote{This was the case for the three boxes (labelled as '\texttt{E1}', `\texttt{W1}' and `\texttt{W2}' in Figs.~\ref{fig:boxes_fit_co10}-\ref{fig:boxes_fit_ci10}) without a S/N$\geq3$ detection of \carbon.}. 
Compared with previous works recovering the same amount of CO flux, our new ${M}^{tot}_{mol}$ estimate is higher than in \cite{Tacconi+99} and \cite{Feruglio+13a}, but consistent with \cite{Papadopoulos+14}.

\subsection{Molecular outflow properties}\label{sec:of_prop}

\begin{table}[tbp]
\centering
\footnotesize
\caption{Summary of source and outflow properties}
\label{table:of_prop}
\begin{tabular}{lc}
\hline
\multicolumn{2}{c}{Source properties:} \\
\hline
$L_{\rm TIR(8-1000\mu m)}$ [erg s$^{-1}$]	&  	$2.71\times10^{45}$$^{(a)}$	   \\
$L_{\rm Bol}$ [erg s$^{-1}$]	 			&  	$3.11\times10^{45}$$^{(b)}$	   \\
$L_{\rm AGN}$ [erg s$^{-1}$] 				&      $(1.1\pm0.4)\times10^{45}$$^{(c)}$ \\
$\alpha_{\rm AGN}\equiv L_{\rm AGN}/L_{\rm Bol}$	&      $0.35\pm0.13$		\\
SFR 	[M$_{\odot}$~yr$^{-1}$] 		        &      $46\pm9$$^{(d)}$			\\
M$_{mol}^{tot}$ [M$_{\odot}$] 		&		$(2.1\pm 0.5)\times10^{10}$$^{(e)}$		\\
\hline
\multicolumn{2}{c}{Molecular outflow properties (estimated in $\S$~\ref{sec:of_prop}):} \\
\hline
$r_{max}$ [kpc]	 &  $2.4\pm0.3$$^\dag$		   \\
$\langle v_{out} \rangle$ [\kms]	& $250\pm50$$^\ddag$	   \\
$\langle \sigma_{out} \rangle$ [\kms]	& $220\pm20$$^\ddag$	   \\
$\langle \tau_{dyn} \rangle$ [Myr] &  $6.5\pm1.8$$^\ddag$   \\
$M_{out}$ [M$_{\odot}$]   & $(1.2\pm0.3)\times 10^{10}$	\\
$\dot{M}_{out}$ [M$_{\odot}$~yr$^{-1}$]  & $2500\pm1200$  \\  
$v\dot{M}_{out}$ [$\rm g~cm~s^{-2}$] &  $(3.1\pm1.2)\times 10^{36}$ \\
$1/2\dot{M}_{out}v^2$ [$\rm erg~s^{-1}$] & $(3.6 \pm 1.6)\times 10^{43}$ \\
\hline
$\eta \equiv \dot{M}_{out}$/SFR   &  $50\pm30$ \\
($v\dot{M}_{out}$)/($L_{AGN}/c$)  & $80\pm50$   \\
($1/2\dot{M}_{out}v^2$)/$L_{AGN}$  & $0.033\pm0.019$  \\
$\tau_{dep} \equiv M_{mol}^{tot}/\dot{M}_{out}$ [Myr] & $8\pm4$ \\   
\hline
\end{tabular}

\begin{flushleft}
\small
$^{(a)}$ From the IRAS Revised Bright Galaxy Sample \citep{Sanders+03}; 
$^{(b)}$ $L_{\rm Bol} = 1.15~L_{\rm TIR}$, following \cite{Veilleux+09};
$^{(c)}$ Total bolometric luminosity of the dual AGN system estimated from X-ray data by \cite{Puccetti+16};
$^{(d)}$ $\rm SFR = (1-\alpha_{AGN})\times 10^{-10}$~$\rm L_{TIR}$, following \cite{Sturm+11}.
$^{(e)}$ Total molecular gas mass in the $12\arcsec \times 6\arcsec$ region encompassing the nucleus and the outflow, derived in $\S$~\ref{sec:of_prop}.\\ 
$^\dag$ Maximum distance at which we detect \carbon~in the outflow at a S/N$>3$, hence the quoted $r_{max}$ should be considered a lower limit constraint allowed by current data. \\
$^\ddag$ Mean values obtained from the analysis of all boxes. \\
\end{flushleft}
\end{table}

In this section we use the results of the spatially-resolved spectral analysis presented in $\S$~\ref{sec:results_boxes} to constrain the mass ($M_{out}$), mass-loss rate ($\dot{M}_{out}$), kinetic power ($1/2\dot{M}_{out}v^2$), and momentum rate ($\dot{M}_{out}v$) of the molecular outflow.   
We first select the boxes in which an outflow component is detected in the \co~spectrum with S/N$\geq3$. As described in $\S$~\ref{sec:results_boxes}, the outflow component is defined as the sum of all Gaussian functions employed by the simultaneous fit that lie outside the rectangular region of the FWHM-$v$ parameter space shown in Fig.~\ref{fig:v_sigma_boxes}.
With this S/N$\geq3$ constraint, 12 boxes (that is, all except \texttt{W1}) are selected to have an outflow component in \co, and for each box
\footnote{All quantities relevant to the individual boxes are identified by an index $i=1, 12$ (e.g. $v_{out,i}$) in order to distinguish them from the corresponding galaxy-integrated quantities (e.g. $v_{out}$).} we measure:

(i) The average outflow velocity ($v_{out,i}$), equal to the mean of the (moduli of the) central velocities of the individual Gaussians classified as `outflow'.

(ii) The molecular gas mass in outflow ($M_{out,i}$), calculated by multiplying the $L^{\prime}_{\co}$ of each outflow component by an appropriate \alphaco. In ten boxes the outflow component is detected with S/N$\geq3$ also in the \carbon~transition, hence for these boxes we can use their corresponding \alphaco~factor (see $\S$~\ref{sec:results_boxes}). For the remaining two boxes (\texttt{E1} and \texttt{W2}), we adopt the galaxy-averaged outflow \alphaco~of $2.1\pm1.2$ (Table~\ref{table:r21_alphaco_values}).  

(iii) The dynamical timescale of the outflow, defined as $\tau_{dyn,i} = R_{i}/v_{out,i}$, where $R_{i}$ is the distance of the centre of the box from RA=16:52:58.900, Dec=02.24.03.950. This definition cannot be applied to the central box (\texttt{C1}) because the so-estimated $R$ would be zero, hence boosting the mass-loss rate to infinite. Therefore, for box \texttt{C1}, we conservatively assume that most of the outflow emission comes from a radius of $1\arcsec$, hence we set $R=0.5$~kpc. For all boxes we assume the uncertainty on $R_{i}$ to be $0.6\arcsec$ ($0.3$~kpc), which is half a beam size. 

(iv) The mass-loss rate $\dot{M}_{out,i}$, equal to $M_{out,i}/\tau_{dyn,i}$. 
All uncertainties are derived by error propagation.

The resulting total outflow mass and mass-loss rate, obtained by adding the contribution from all boxes, are respectively
$M_{out} = (1.2\pm0.3)\times 10^{10}~M_{\odot}$ and $dM_{out}/dt = 2500\pm1200$~$M_{\odot}~yr^{-1}$. As discussed in Appendix~\ref{sec:boxspectra_appendix}, the largest contribution to both $\dot{M}_{out}$ and its uncertainty is given by the central box. Indeed, box \texttt{C1} has at the same time the highest estimated $M_{out}$ and the smallest - and most uncertain - $R$, because the outflow is launched from within this region, likely close to the mid-point between the two AGNs as suggested by Fig.~\ref{fig:outflow_maps}(d,e). 

Similar to the mass-loss rate, we calculate the total kinetic power and momentum rate of the outflow by summing the contribution from all boxes with a \co~outflow component, and we obtain respectively: $1/2\dot{M}_{out}v_{out}^2\equiv \sum_i 1/2\dot{M}_{out,i}v_{out,i}^2 = (0.033\pm0.019)~L_{AGN}$ and $v \dot{M}_{out} \equiv \sum_i v_i\dot{M}_{out,i} = (80\pm50)~L_{AGN}/c$. 
If all the gas carried by the outflow escaped the system and the mass-loss continued at the current rate, the depletion time-scale of the molecular gas reservoir in NGC~6240 would be $\tau_{dep} = 8\pm4$~Myr. 
All the relevant numbers describing the properties of the source and of the molecular outflow are reported in Table~\ref{table:of_prop} and will be discussed in $\S$~\ref{sec:disc_feedback} in the context of feedback models.

Very stringent lower limits on the outflow energetics can be derived by assuming that its \co~emission is fully optically thin. For optically thin gas and $T_{ex}=30$~K, the \alphaco~factor would be $\sim0.34$ \citep{Bolatto+13}, and the outflow mass and mass-loss rate would be $M_{out} = (1.98\pm0.09)\times10^9~ M_{\odot}$ and $dM_{out}/dt = 430 \pm 160$~$M_{\odot}~yr^{-1}$. We however stress that the assumption of fully optically thin \co~emission in the outflow is not supported by our data, which instead favour an \alphaco~factor for outflowing gas that is intermediate between the optically thin and the optically thick values (for solar metallicities).

\subsection{Physical properties of quiescent and outflowing gas}\label{sec:alphaco_r21}

\begin{figure*}[tb]
\centering
\includegraphics[clip=true,trim=6cm 4.4cm 2.5cm 3.7cm, width=.8\columnwidth,angle=180]{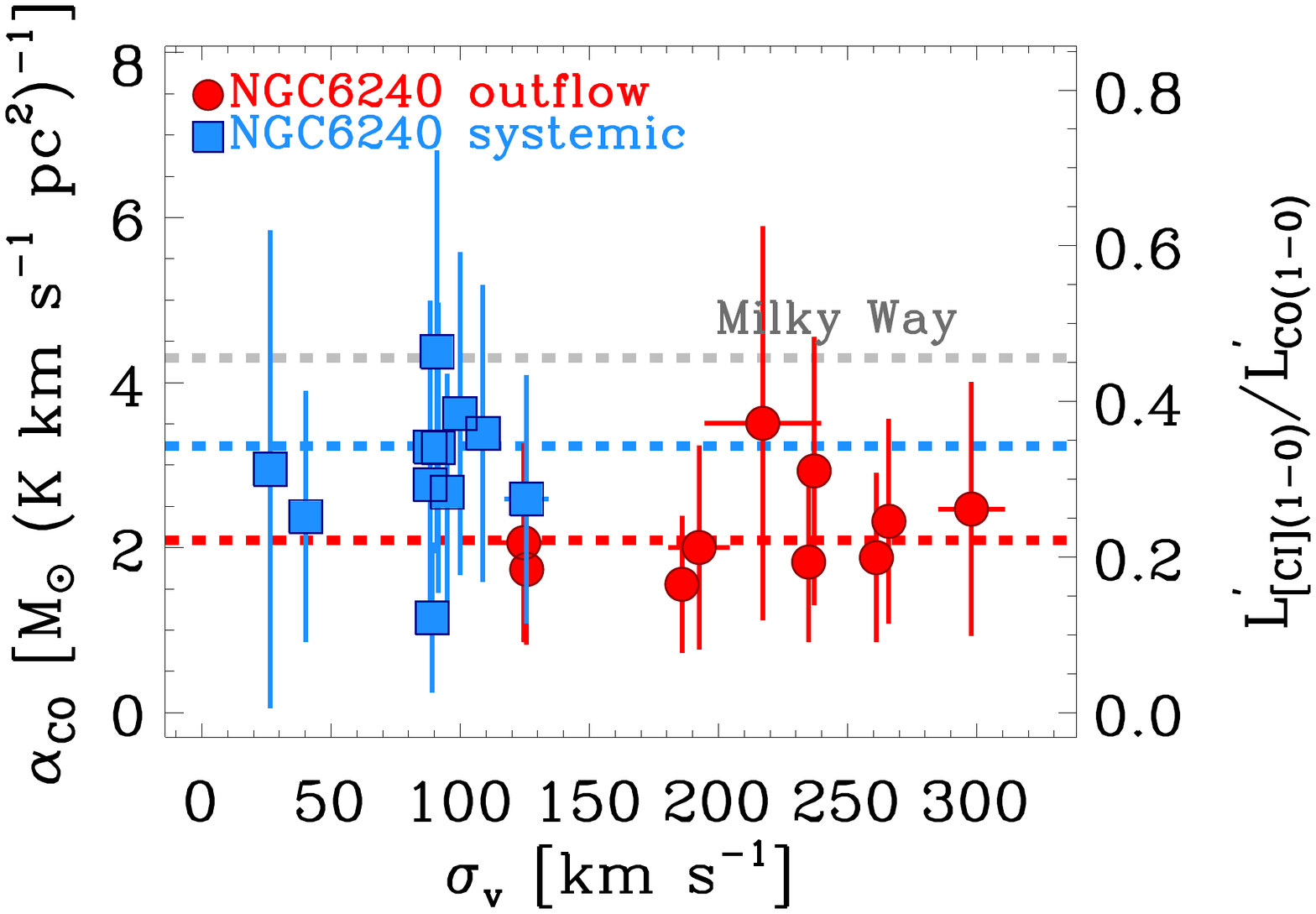}\quad
 \includegraphics[clip=true,trim=6cm 4.4cm 2.5cm 3.7cm, width=.8\columnwidth,angle=180]{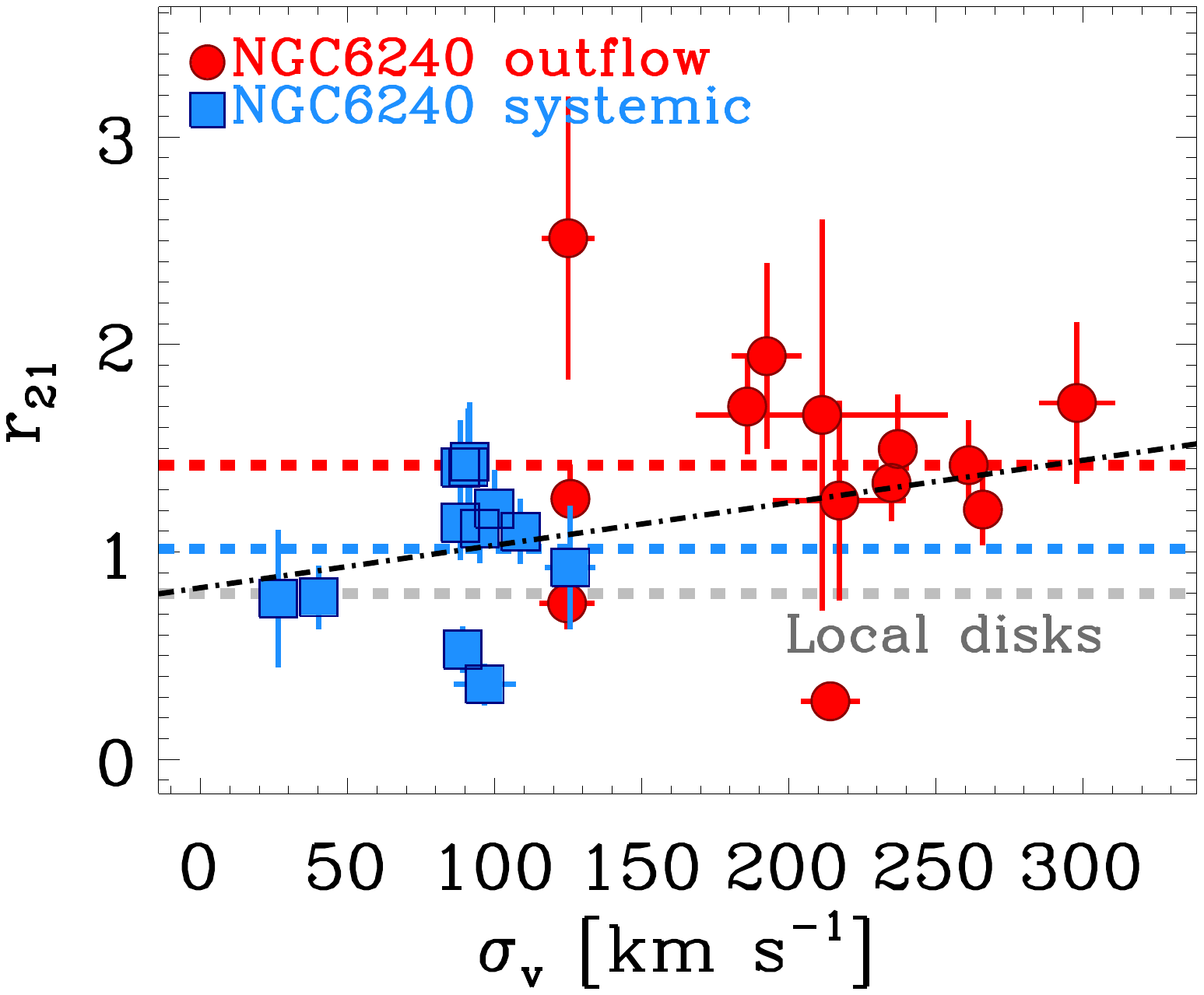}\\
    \includegraphics[clip=true,trim=6cm 4.4cm 2.5cm 3.7cm, width=.8\columnwidth,angle=180]{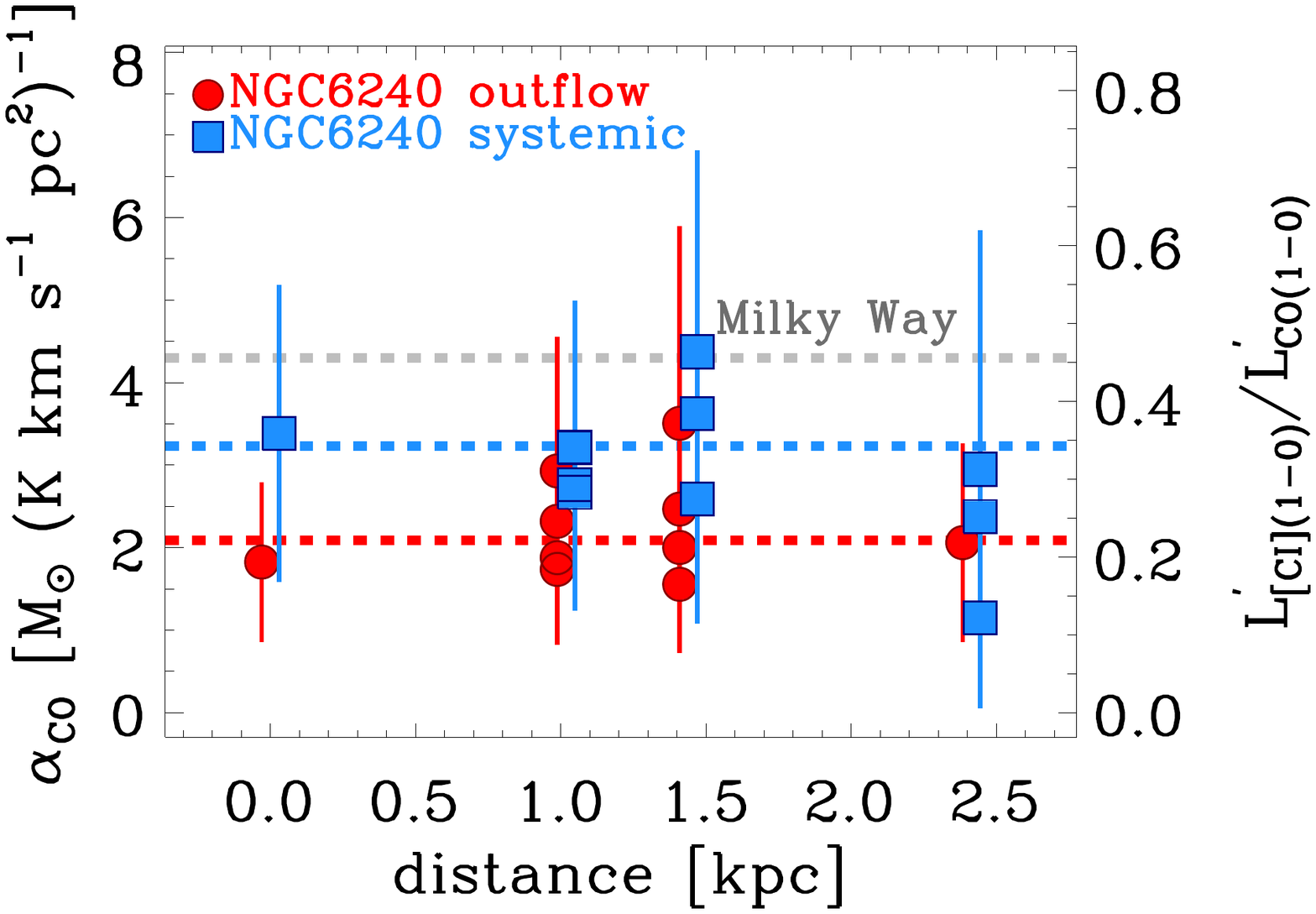}\quad
  \includegraphics[clip=true,trim=6cm 4.4cm 2.5cm 3.7cm, width=.8\columnwidth,angle=180]{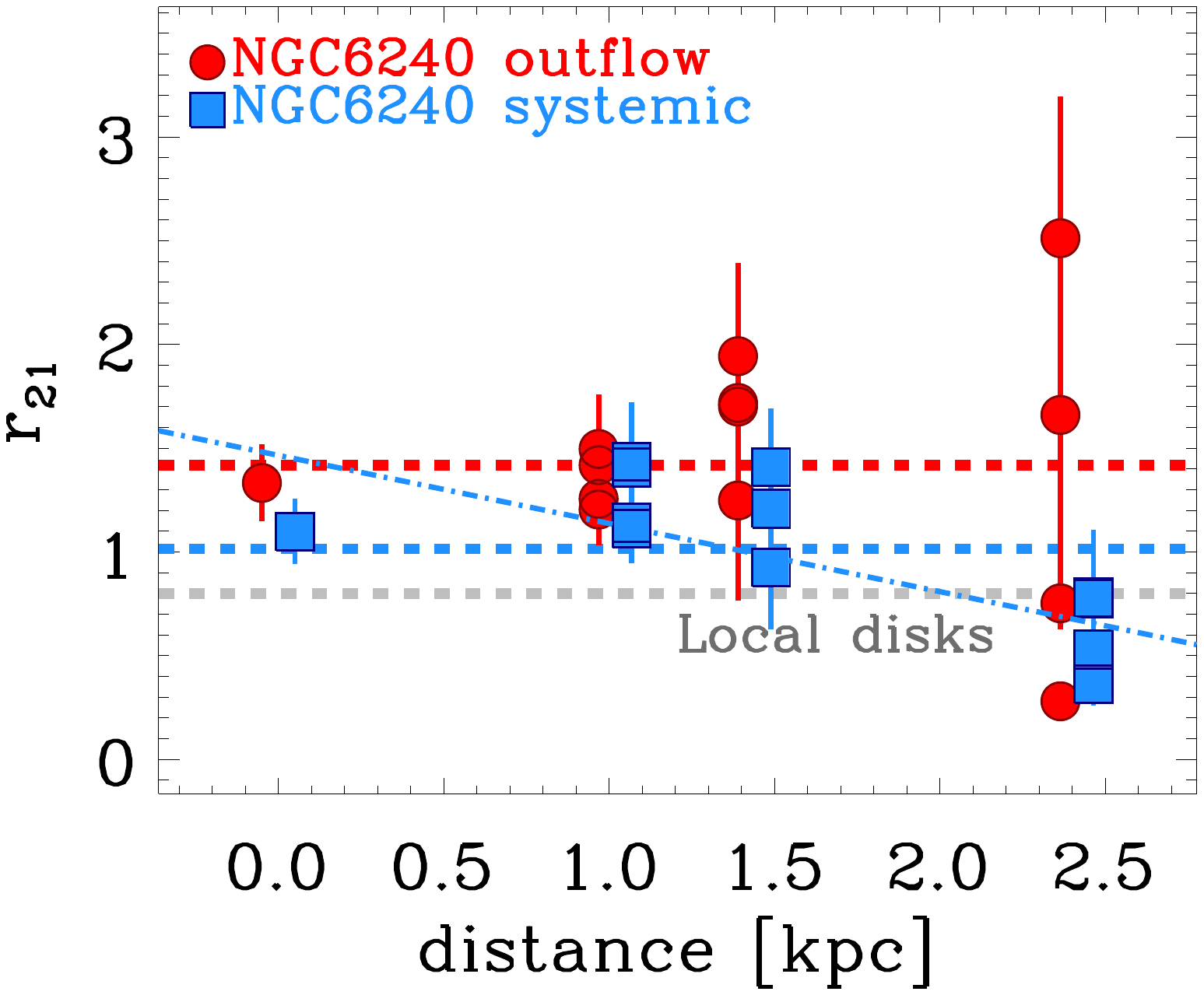}\\
\caption{\alphaco~({\it left}) and $r_{21}$ ({\it right}) as a function of the average velocity dispersion ({\it top}) and of the distance from the nucleus ({\it bottom}) of the corresponding molecular line components. A detailed explanation on how  \alphaco~and $r_{21}$ were calculated can be found in $\S$~\ref{sec:results_boxes}. The $y$ axis on the right side of the \alphaco~plots shows the corresponding \carbon/\co~line luminosity ratio.
The horizontal blue and red dashed lines are the mean values reported in Table~\ref{table:r21_alphaco_values} for the systemic and outflow components, respectively. The grey lines indicate the Milky Way \alphaco~factor (\citealt{Bolatto+13}, {\it left panels}) and the average $r_{21}=0.8$ measured in star forming galaxies (\citealt{Leroy+09}, {\it right panels}). 
The best-fits obtained from a Bayesian linear regression analysis following the method by \cite{Kelly07} are plotted using dot-dashed lines: black lines show the best fits to the total sample, whereas blue and red lines correspond to the fits performed separately on the systemic and outflowing components.}
\label{fig:plots}
\end{figure*}

Using the results of the spatially-resolved analysis presented in $\S$~\ref{sec:results_boxes}, we now study how the \alphaco~and $r_{21}$ parameters vary as a function of velocity dispersion ($\sigma_v$) and projected distance ($d$) from the nucleus of NGC~6240. The relevant plots are shown in Figure~\ref{fig:plots}. To investigate possible statistical correlations, we conduct a Bayesian linear regression analysis of the relations in Fig.~\ref{fig:plots}
following \cite{Kelly07}\footnote{We used the IDL routine \texttt{linmix\_err.pro}}. 

The left panels of Fig.~\ref{fig:plots} do not indicate any statistically significant relation between \alphaco~and either $\sigma_v$ or $d$. Instead, they show that the \alphaco~factor is systematically higher - although formally only at a significance of $1.2\sigma$ (Table~\ref{table:r21_alphaco_values}) - in the quiescent gas than in the outflow, regardless of the velocity dispersion of the clouds, or of their position with respect to the merger nucleus.
For the {\it non-outflowing} components, the \alphaco~factors are at least twice the so-called (U)LIRG value \citep{Downes+Solomon98}, and reach up to Galactic values.
This result is consistent with the multi-transition analysis by \cite{Papadopoulos+14}, and is likely due to the state of the dense gas phase that low-$J$ CO lines alone cannot constrain, but which instead is accounted for when using \carbon~as a molecular mass tracer. 
Nevertheless the {\it outflowing $\rm H_2$} gas has lower \alphaco~values than the quiescent ISM. This is indeed expected from the ISM physics behind \alphaco~for warm and strongly unbound gas states \citep{Papadopoulos+12_ApJ}, i.e. the type of gas that we expect to be embedded in outflows. In particular,
in the case that molecular outflows are ubiquitous in (U)LIRGs as suggested by observations \citep{Sturm+11,Veilleux+13, Spoon+13,Cicone+14}, {\it the outflow may be the location of the diffuse and warm molecular gas phase that is not contained in self-gravitating cooler clouds - a sort of `intercloud' medium advocated by some of the previous analyses based solely on low-J CO, $^{13}$CO line observations} \citep{Aalto+95,Downes+Solomon98}. 
Furthermore, the flat trend between \alphaco~and $d$ observed in Fig.~\ref{fig:plots} does not support the hypothesis that the lower \alphaco~values in (U)LIRGs are related to the collision of the progenitors' disks, since in this case we would naively expect the lower \alphaco~clouds to be concentrated in the central regions of the merger. 
The \alphaco~values measured for the outflow components are however significantly higher than the optically thin value, suggesting that not all of the outflowing material is diffuse and warm, but there may still be a significant amount of dense gas. These results are further discussed and contextualised in
 $\S$~\ref{sec:disc_outflowrole}.

The right panels of Fig.~\ref{fig:plots} show a weak correlation between the $r_{21}$ and $\sigma_v$ (correlation coefficient, $\rho=0.4\pm0.2$) and an anti-correlation with the distance, although only for the systemic/quiescent  components ($\rho=-0.7\pm0.2$). The corresponding best fit relations, of the form $r_{21}=\alpha + \beta x$, plotted in Fig.~\ref{fig:plots}, have $(\alpha,\beta)=(0.8\pm0.2, 2.1\pm1.4 \times10^{-3})$ for $x=\sigma_v$, and $(\alpha,\beta)=(1.5\pm0.2, -0.33\pm0.13)$ for $x=d$. 
The systemic ISM shows $0.8\lesssim r_{21}\lesssim1.4$, whereas the outflow is characterised by higher ratios, with most components in the range $1.2\lesssim r_{21}\lesssim2.5$, although we observe a large spread in $r_{21}$ values at $d>2$~kpc. 

\cotwo/\co~luminosity ratios of $r_{21}\sim0.8-1.0$ are typically found in the molecular disks of normal spiral galaxies \citep{Leroy+09} and are indicative of optically thick CO emission with $T_{kin}\sim10-30$~K (under LTE assumptions). 
Nevertheless such low-$J$ CO line ratios, in absence of additional transitions, are well-known to be highly degenerate tracers of the average gas physical conditions. Higher-$J$ data of CO, molecules with larger dipole moment, and isotopologues can break such degeneracies. Such studies exist for NGC~6240 \citep{Greve+09,Meijerink+13,Papadopoulos+14}, and found extraordinary states for the molecular gas, with average densities typically above $\rm 10^{4}~cm^{-3}$ and temperatures $T_{kin} \sim 30-100$~K.

On the contrary, global \cotwo/\co~ratios exceeding unity have a lower degree of degeneracy in terms of the extraordinary conditions that they imply for molecular gas, as they require warmer ($T_{kin}\gtrsim100$~K) and/or strongly unbound states \citep{Papadopoulos+12_MNRAS}. 
In NGC~6240, optical depth effects are most likely at the origin of the $r_{21}>1$ values. More specifically, such ratios can result from highly non-virial motions (e.g. the large velocity gradients of the outflowing clouds), causing the CO lines to become partially transparent, as also supported by the tentative trend of increasing $r_{21}$ with $\sigma_v$ (Figure~\ref{fig:plots}). This finding independently strengthens our explanation for the lower \alphaco~factors derived for the outflowing gas, which are intermediate between an optically thin and an optically thick value (for typical solar CO abundances). 

\section{Discussion}\label{sec:discussion}

\subsection{Assumptions and caveats of our analysis}\label{sec:disc_caveats}
Our results build, on the one hand, on the identification of the outflow components, and on the other hand on the assumption that \co~and \carbon~trace the same molecular gas, implying that $M_{mol}$ can be measured from \carbon. In this section we further comment on these steps and discuss their caveats and limitations.

\subsubsection{The outflow identification}\label{sec:disc_caveats_outflowid}
The outflow identification is a fundamental step of our analysis, and leads to one of the most surprising findings, i.e.
that $60\pm20$\% of the molecular ISM in NGC~6240 belongs to the outflow. This unprecedented result may hold the key to finally 
understanding the extreme ISM of this source, which makes it an outlier even compared to other (U)LIRGs, as acknowledged by several authors \citep{Meijerink+13, Papadopoulos+14, Israel+15}. 
For example, \cite{Meijerink+13} suggested that the CO line emission in NGC~6240 is dominated by gas settling down after shocks, which would
be consistent with gas cooling out of an outflow. A massive outflow would also explain why the gaseous and stellar kinematics are 
decoupled \citep{Engel+10, Tacconi+99}. 

In $\S$~\ref{sec:results_boxes} we have ascribed to the outflow all spectral line components with 
FWHM$>400$~\kms, $v<-200$~\kms {\it or} $v>+250$~\kms detected within the central $12\arcsec\times6\arcsec$ region investigated in this paper. However, the spatial information is also crucial for identifying outflowing gas,
especially in a source undergoing a major merger, since the outflow signatures may be degenerate with gravity-driven dynamical effects. In the specific case of NGC~6240, as explained below and shown in detail in Appendix~\ref{sec:boxspectra_appendix}, the high S/N and spatial resolution of our observations allow us to disentangle feedback-related effects from other mechanisms and reliably identify the outflow emission.

During a galaxy collision, high-$v$/high-$\sigma_v$ gas can be concentrated in the nuclear region as a consequence of gravitational torques, which cause a fraction of the gas to lose angular momentum and flow toward the center. At the same time, gravitational torques and tidal forces can drive out part of the gas from the progenitors' disks and form large-scale filaments denominated `tidal tails' and `bridges'. However, in the case of NGC~6240, these gravity-induced mechanisms can hardly explain the kinematics and morphology of the $\sim10$~kpc-scale, wide opening angle-emission shown in Figs.~\ref{fig:outflow_maps}-\ref{fig:CI10_moments}. In particular, the high-$v$/high-$\sigma_v$ structures revealed by the \carbon~moment maps, which are correlated with features observed on much larger scales (see $\S$~\ref{sec:results_outflowmap}), cannot be due to nuclear inflows. In this case, we would indeed expect the $\sigma_v$ of the gas to be enhanced toward the nucleus (or nuclei), rather than in offset positions that are several 100s of pc away from the nuclei or from the geometric center of the AGN pair (see for example the different signature of outflows and inflows in the velocity dispersion maps shown by \cite{Davies+14}). The hourglass-shaped configuration visibile in the \carbon~velocity dispersion map is more suggestive of an outflow opening toward east and west, i.e. along the same directions of expansion of the high-$v$ gas.

On larger scales, tidal tails or bridges produced in galaxy collisions may also affect the dynamical state of the ISM. However, the line-widths of the molecular emission from such filamentary structures are rather low ($\sim 50-100$~\kms, \citealt{Braine+01}). Therefore, in order to reproduce the spatially- and kinematically- coherent structure shown in Figs.~\ref{fig:outflow_maps}, and especially the 
spatial overlap across several kpc between the highly blue-shifted and red-shifted emissions (Figure~\ref{fig:CO21wings_overlap}), one would need to postulate a very specific geometry where several tidal tails overlap along the line of sight across more than 10 kpc.

Based on these considerations, and on the detailed discussion reported in Appendix~\ref{sec:boxspectra_appendix}, we conclude that
other mechanisms such as rotating disks or gravity-induced dynamical motions, possibly also coexisting in NGC~6240, are unlikely to significantly affect our outflow energetics estimates.

\subsubsection{Combining \carbon~and \co~data to infer \alphaco~and $M_{mol}$}\label{sec:disc_caveats_alphaco}

The second key step of our analysis is to combine the \carbon~and \co~line observations to derive molecular gas masses. 
As described in $\S$~\ref{sec:results_total}-\ref{sec:of_prop}, our strategy is to use the places where \carbon~and \co~are both detected at a S/N$\geq3$ to measure
the corresponding \alphaco. Molecular gas masses are then computed by using the \co~data. In particular, we select components where \co~is detected at a S/N$\geq3$ and convert $L^{\prime}_{\co}$ into $M_{mol}$ by employing either the corresponding [CI]-derived \alphaco (possible only if \carbon~is also detected with S/N$\geq3$) or alternatively by using the mean \alphaco~value appropriate for that component (i.e. `global', `systemic', or `outflow', Table~\ref{table:r21_alphaco_values}). 

The fundamental underlying assumption is that \carbon~and \co~trace the same material. Earlier theoretical works envisioned neutral Carbon to be confined in the external (low extinction $A_V$) layers of molecular clouds, hence to probe a different volume compared to CO. However, as
discussed by \cite{Papadopoulos+04}, this theory was dismantled by observations finding a very good correlation between [CI] and CO as well as uniform [CI]/CO ratios across a wide range of Galactic environments, including regions shielded from FUV photons (e.g. \citealt{Keene+85, Ojha+01, Tanaka+11}). The few available observations of [CI] lines in local galaxies have further supported the concurrence of CO and [CI] in different physical conditions \citep{Israel+15, Krips+16}.

The good mixing of CO and [CI] could be a consequence of turbulence and/or cosmic rays. Turbulent diffusion can merge any [CI]-rich H$_2$ phase (expected to prevail in low $A_V$ regions) with the more internal CO-rich H$_2$ gas, hence uniforming the [CI]/CO abundance ratio throughout molecular clouds \citep{Glover+15}. Cosmic rays, by penetrating deep into molecular clouds and so destroying CO (but not H$_2$) over larger volumes compared to FUV photons, can also help enrich the internal regions of clouds with neutral Carbon \citep{Bisbas+15, Bisbas+17}. Both mechanisms are expected to be efficient in (U)LIRGs and in their molecular outflows. The latter are (by definition) highly turbulent environments. Furthermore, cosmic rays originating in the starburst nuclei can leak along such outflows hence influencing the chemistry of their embedded ISM (see discussion in \cite{Papadopoulos+18}, and recent results by \cite{Gonzalez-Alfonso+18}). For these reasons, we can assume that CO and [CI] trace the same molecular gas, for both the quiescent and outflowing components of NGC~6240.

Thanks to the simple three-level partition function of neutral Carbon, and to its lines being optically thin in most cases (including NGC~6240, \cite{Israel+15}), the main sources of uncertainties for [CI]-based mass estimates are $X_{\rm CI}$ and $T_{ex}$ (Eq~\ref{eq:molmass}). 
Previous observations indicate very little variations in $X_{\rm CI}$ in the metal-enriched ISM of IR-luminous galaxies at different redshifts \citep{Weiss+03,Weiss+05,Danielson+11,Alaghband-Zadeh+13}, including the extended ($r>10$~kpc) circum-galactic medium of the Spiderweb galaxy \citep{Emonts+18}. In our calculations we assumed $X_{\rm CI} = (3.0\pm1.5)\times 10^{-5}$ to take 
into account a systematic uncertainty associated with the [CI]/H$_2$ abundance ratio.
Because of the particular LTE partition function of neutral Carbon, [CI]-derived masses depend little on $T_{ex}$ for $T_{ex}\gtrsim15$~K. 
We set $T_{ex}=30$~K, which is consistent with the value that can be estimated from 
the global [CI]2-1/1-0 brightness temperature ratio measured in NGC~6240 \citep{Papadopoulos+14}. 

Therefore, our assumptions regarding the conversion between \carbon~line data and $M_{mol}$ are well justified by previous results. 
However, we caution that a giant galactic-scale outflow such as the one hosted by NGC~6240 constitutes an unprecedented environment for molecular gas clouds, and there is no comparable laboratory in our Galaxy that can be used as a reliable reference. The study of the physical conditions of such outflows has only just started, and this is the first time that the \carbon~line emission from high-velocity gas components extending by several kpc has been imaged at high spatial resolution. Further investigation is needed, and our work constitutes just a starting point.

\subsection{The role of outflows in the global \alphaco~factor}\label{sec:disc_outflowrole}

The average \alphaco~factors that we measure for the quiescent and outflowing components of the ISM in NGC~6240 (Table~\ref{table:r21_alphaco_values}) are both higher than the classic (U)LIRG \alphaco. How can we reconcile this result with previous works advocating for significantly lower \alphaco~values in (U)LIRGs? In the case of NGC~6240, our analysis has highlighted several effects that may have plagued previous \alphaco~estimates:
\begin{enumerate}
\item The widespread presence of outflowing gas implies that, at any location within this merger, the molecular line emission includes a 
significant contribution from the outflow, with its overall lower \alphaco~and 
higher $r_{21}$. As a result, an analysis of the global ISM conditions (especially if based only on low-$J$ CO lines, see also point 3 below) would get contaminated by the warm unbound H$_2$ envelopes in the outflow, and their larger $L^{\prime}_{CO}/M_{H2}$ ratios would drive down the global \alphaco~estimate \citep{Yao+03, Papadopoulos+12_ApJ}.
\item The outflow dominates the velocity field of the H$_2$ gas throughout the entire source, including the central region (Figure~\ref{fig:CI10_moments}). The apparent nuclear north-south velocity gradient identified in previous CO line data
\citep{Tacconi+99, Bryant+Scoville99} is actually not compatible with ordered rotation once observed at higher spatial resolution, but 
it shows several features distinctive of the outflow. Therefore, the assumption that the molecular gas in this area is dominated by ordered motions is broken, making any dynamical mass estimate unreliable (if the outflow is not properly taken into account).
\item Previous analyses based only on low-$J$ CO lines have probably missed a substantial fraction of the denser gas phase that is instead accounted
for when using the optically thin \carbon~line as a gas mass tracer, or when probing the excitation of the ISM using high-$J$ CO transitions and
high density molecular gas tracers. 
Indeed, the \alphaco~value derived for the quiescent gas reservoir is consistent with a significant contribution from a dense gas state.
Even in the outflow, the average \alphaco~is still significantly higher than the optically thin value,
hence the presence of dense gas may not be negligible. A conspicuous dense gas phase has already been demonstrated in a few other galaxy-wide molecular outflows \citep{Aalto+12, Sakamoto+14, Garcia-Burillo+14, Alatalo+15}, and its presence would make more likely the formation of stars within these outflows \citep{Maiolino+17}. 
\item The molecular gas emission in NGC~6240 is clearly
very extended - both spectrally and spatially. As a result, at least some of the previous interferometric observations (especially `pre ALMA') may have been severely affected by incomplete {\it uv} coverages filtering out the emission on larger scales, hence impacting on the measured line fluxes and sizes. Furthermore, as already noted by \cite{Tacconi+99}, an insufficient spectral bandwidth may have hindered a correct baseline and/or continuum fitting and subtraction. The latter can be an issue for both single dish and interferometric observations, including observations with ALMA if only one spectral window is employed to sample the line.
\end{enumerate}
Since molecular outflows are a common phenomenon in local (U)LIRGs \citep{Sturm+11,Veilleux+13,Spoon+13,Cicone+14,Fluetsch+18}, at least some of the above considerations may be generalised to their entire class. {\it Therefore, it is possible that the so-called (U)LIRG \alphaco~factor is an artefact resulting from modelling the molecular ISM of such sources containing massive H$_2$ outflows}.  

\subsection{An interplay of feedback mechanisms at work}\label{sec:disc_feedback}

The extreme spatial extent of its molecular outflow, makes NGC~6240 one of the few sources - all powerful quasars - hosting H$_2$ outflows with sizes of $\gtrsim10$~kpc (\cite{Veilleux+17}, see also the 30~kpc-size [CII]$\lambda158\mu m$ outflow at $z=6.4$ studied by \cite{Cicone+15}). In comparison, the H$_2$ gas entrained in the well-studied starburst-driven winds of M~82 and NGC~253 reaches at maximum scales of $\sim1-2$~kpc \citep{Walter+02,Walter+17}. Furthermore, among all the large-scale molecular outflows discovered so far in quasar host galaxies, the outflow of NGC~6240 is the one
that has been observed at the highest spatial resolution ($\sim120$~pc). Indeed, the ALMA \carbon~line data allowed us to probe deep into the nuclear region of the merger, close to the launching point of the molecular wind, and surprisingly revealed that the outflow emission peaks between the two AGNs rather than on either of the two. This is apparently at odds with an AGN radiative-mode feedback scenario, in which the multiphase outflow is expected to be generated close to the central engine \citep{Costa+15}. 

Nevertheless, the role of the AGN(s) is certified by the extreme energetics of the molecular outflow, which has been constrained here with unprecedented accuracy. By comparing our $M_{out}$ and $M_{mol}^{tot}$ estimates (Table~\ref{table:of_prop}), it appears that $60\pm20$~\% of the molecular medium is involved in the outflow. The estimated mass-loss rate of $2500\pm1200$~$M_{\odot}~yr^{-1}$ corresponds to $\rm \eta \equiv \dot{M}_{out}/SFR = 50\pm30$, whereas the lower limit on $\rm \dot{M}_{out}$, calculated 
using the optically thin \alphaco~prescription, corresponds to $\rm \eta \equiv \dot{M}_{out}/SFR = 9\pm4$. Such high mass loading factors are inconsistent with a purely star formation-driven wind. As a matter of fact, stellar feedback alone can hardly bear outflows with $\eta$ much higher than unity. Cosmological hydrodynamical simulations incorporating realistic stellar feedback physics, by including mechanisms other than supernovae, can reach up to $\eta\sim10$ \citep{Hopkins+12}. However, because $\eta$ in these simulations anti-correlates with the mass of the galaxy, the highest $\eta$ values are generally predicted for dwarf galaxies, whereas for galaxies with baryonic masses of several 10$^{10}~M_{\odot}$ such as NGC~6240, the $\eta$ achievable by stellar feedback can be at maximum $2-3$. 

Based on its energetics, we can therefore rule out that the massive molecular outflow observed in NGC~6240 is the result of star formation alone. The outflow energetics can instead be fully accommodated within the predictions of AGN feedback models \citep{Faucher-Giguere+12, Zubovas+King14, Costa+14}. In addition, these models can explain the multi-wavelength properties of NGC~6240. At optical wavelengths, NGC~6240 is known to host a ionised wind \citep{Heckman+90}, with large-scale superbubbles expanding by tens of kpc towards north-west and south-east \citep{Veilleux+03,Yoshida+16}. The H$\alpha$ emission from the ionised wind shows a close spatial correspondence with the soft X-ray continuum, suggesting the presence of gas cooling out of a shocked medium \citep{Nardini+13}. Furthermore,
\cite{Wang+14} detected a diffuse component in hard-X-ray continuum and FeXXV line emission, 
tracing $T\sim7\times10^7$~K gas between the two AGNs (north-west of the southern nucleus, similar to the \carbon~blue wing in Fig~\ref{fig:outflow_maps}d), as well as in kpc-scale structures that are remarkably coincident with both the strong NIR H$_2$ emission \citep{Max+05, vanderWerf+93} and the H$\alpha$ filaments. 

At radio wavelengths, \cite{Colbert+94} reported the detection of non-thermal continuum emission with a steep spectrum extending by several kpc in an arc-like structure west of the AGNs, later confirmed also by \cite{Baan+07}, together with a possibly similar feature on the eastern side. This structure lacks a clear spatial correspondence with optical or NIR starlight (which excludes a starburst origin) and its complex morphology suggests a connection with the H$\alpha$ outflow. Theoretically, the association between an AGN-driven wind and extended non-thermal radiation (due to relativistic electrons accelerated by the forward shock) has been predicted by \cite{Nims+15}. Observationally, the rough alignment of the western arc-like structure discovered by \cite{Colbert+94} with the molecular outflow studied in this work
would also support this hypothesis, although the current data do not allow us to probe the presence of H$_2$ outflowing gas at the exact position of the radio emission. The total radio power of this arclike feature is comparable with that of extended radio structures previously observed in radio-quiet AGNs with prominent outflows \citep{Morganti+16}. Future facilities like the Cherenkov Telescope Array (CTA) may reveal the $\gamma$-ray counterpart of the non-thermal emission, as expected for an AGN outflow shock \citep{Lamastra+17}.

In summary, all these multi-wavelength observational evidences point to a radiative-mode AGN feedback mechanism \citep{Faucher-Giguere+12, Nims+15}. However, at the same time, it is difficult to reconcile a classic model with an H$_2$ outflow whose emission does not peak on either of the two AGNs. A complex interplay of stellar and AGN feedback must be at work in this source (see also \citealt{Muller-Sanchez+18}), and we cannot exclude the additional contribution from compact radio-jets \citep{Gallimore+Beswick04}, which may be accelerating part of the cold material \citep{Mukherjee+16}.
Finally, positive feedback may also be at work in NGC~6240. There is indeed a striking correspondence between (i) the morphology of the approaching side of the outflow north-west of the southern AGN (Fig.~\ref{fig:outflow_maps}d), (ii) a peak of dust extinction, and (iii) a stellar population with unusually large stellar $\sigma_v$ and blue-shifted velocities \citep{Engel+10}. The latter, according to \cite{Engel+10}, may have been formed recently as a result of crushing of molecular clouds, which could be related to the observed outflow event \citep{Maiolino+17, Zubovas+King14} provided these stars are not older than a few Myr.

\section{Summary and conclusions}\label{sec:conclusion}

A powerful multiphase outflow shapes the distribution of gas in NGC~6240, and it is likely at the origin of many of the extraordinary features that 
for years have puzzled scientists studying this source. 
In this work we used new ALMA \carbon~line observations, in combination with ALMA \cotwo~and IRAM PdBI \co~line data, to study the morphology, energetics, and physical state of the molecular component of the outflow. Our main findings are:
\begin{itemize}
\item The molecular outflow extends by more than 10~kpc along the east-west direction, and it is clearly detected in both its approaching (blue-shifted) and receding (red-shifted) sides. Its emission peaks between the two AGNs, rather than on either of the two. Furthermore, the outflow dominates 
the H$_2$ gas velocity field in the merger nucleus, as shown by the presence of a striking hourglass-shaped feature in the high-res ($\sim0.24\arcsec$) \carbon~line moment~2 map. This high-$\sigma_v$ structure, aligned east-west, traces the launch base of the kpc-scale outflow.
The outflow, with its large flux contribution to the molecular line emission in the nucleus, can explain both the high gas turbulence and the strong decoupling of stellar and gaseous kinematics evidenced in this source by previous works.
\item We combined the \carbon~and \co~line observations to derive the \alphaco~factor in the outflow, which is on average $\langle\alpha_{CO}\rangle=2.1\pm1.2~\rm M_{\odot} (K~km~s^{-1}~pc^2)^{-1}$. The information on the \alphaco, in conjunction with a spatially-resolved spectral analysis of the molecular line emission, allowed us to constrain with unprecedented accuracy the energetics of the molecular outflow. We estimate that the outflow entrains $M_{out}=(1.2\pm0.3)\times10^{10}~M_{\odot}$, corresponding to 
$60\pm20$~\% of the molecular reservoir of NGC~6240. The total mass-loss rate is $\dot{M}_{out} = 2500\pm1200~M_{\odot}~yr^{-1} = 50\pm30$~SFR, which energetically rules out a solely star formation-driven wind. 
\item For the quiescent gas components, the \alphaco~factors are on average higher than in the outflow (irrespective of their distance from the nucleus), with a mean value of 
$\langle\alpha_{CO}\rangle=3.2\pm1.8~\rm M_{\odot} (K~km~s^{-1}~pc^2)^{-1}$, i.e. 
at least twice the so-called (U)LIRG value. This result is consistent with recent multi-transition ISM analyses and is likely due to a
dense gas phase that cannot be constrained by using low-J CO lines alone, but which is instead accounted for when using \carbon~as a molecular gas tracer.
\item We observe a tentative trend of increasing $r_{21}$ ratios with $\sigma_v$ and measure $r_{21}>1$ values in the outflow ($\langle r_{21}\rangle=1.4\pm0.3$), while $r_{21}\simeq1$ for quiescent gas. We explain the $r_{21}>1$ ratios with optical depth effects, whereby the highly non virial motions of the outflowing clouds cause the CO lines to become partially transparent.
\item Based on the finding that lower \alphaco~and higher $r_{21}$ values are typical of the outflowing clouds, we propose that molecular outflows are the location of the warm and strongly unbound phase - the `intercloud medium' invoked by previous studies - that drives down the global \alphaco~in (U)LIRGs. However, we note that the [CI]-based \alphaco~factor derived for the outflow is higher than the optically thin value, suggesting that not all of the outflowing material is in such warm diffuse phase but that there may still be a significant amount of dense gas entrained.
\item The outflow kinetic power and momentum rate, respectively equal to $(0.033\pm0.019)~L_{AGN}$ and $(80\pm50)~L_{AGN}/c$, could be fully accommodated within the predictions of AGN `blast-wave' feedback models. However, the puzzling outflow morphology, with a launch region situated between the two AGNs, and a direction of expansion perpendicular to the axis connecting the two nuclei, challenges a classic AGN feedback scenario. A complex interplay of stellar and AGN feedback processes must be at work in NGC~6240.
\end{itemize}

\acknowledgments

This project has received funding from the European Union's Horizon 2020 research and innovation programme under the Marie Sk\l{}odowska-Curie grant agreement No 664931. The research leading to these results has received funding from the European Union's Horizon 2020 research and innovation programme under grant agreement No 730562 [RadioNet]. 
R.M. acknowledges ERC Advanced Grant 695671 `QUENCH' and support by the Science and Technology Facilities Council (STFC). E.T. acknowledges support from FONDECYT regular grant 1160999 and Basal-CATA PFB-06/2007.
G.C.P. acknowledges support from the University of Florida. We thank the referee for his/her constructive report, which helped us improve the discussion of the results.
This paper makes use of the following ALMA data: 
ADS/JAO.ALMA \#2015.1.00717.S and \#2015.1.00370.S.
ALMA is a partnership of ESO (representing its member states), NSF (USA) and NINS (Japan), together with NRC (Canada), MOST and ASIAA (Taiwan), and KASI (Republic of Korea), in cooperation with the Republic of Chile. The Joint ALMA Observatory is operated by ESO, AUI/NRAO and NAOJ.
This publication makes use of observations carried out with the IRAM Plateau de Bure Interferometer. IRAM is supported by INSU/CNRS (France), MPG (Germany) and IGN (Spain).
C.C. thanks Sandra Burkutean for helping her with the combination of the ALMA and ACA B8 datacubes and Alvaro Hacar Gonzalez for suggesting to use the ACA B8 data as a source model in the cleaning of the ALMA B8 datacubes, which significantly improved the results. 

%

\vspace{5mm}
\facilities{ALMA, IRAM PdBI}


\software{CASA v4.6.0, GILDAS\citep{Pety05}}



\bibliography{outflow_bib}

\appendix

\section{Additional \co~and~\cotwo~outflow maps}\label{sec:additional_COmaps}

\begin{figure*}[tbp]  
\centering
\includegraphics[width=.6\textwidth,angle=90]{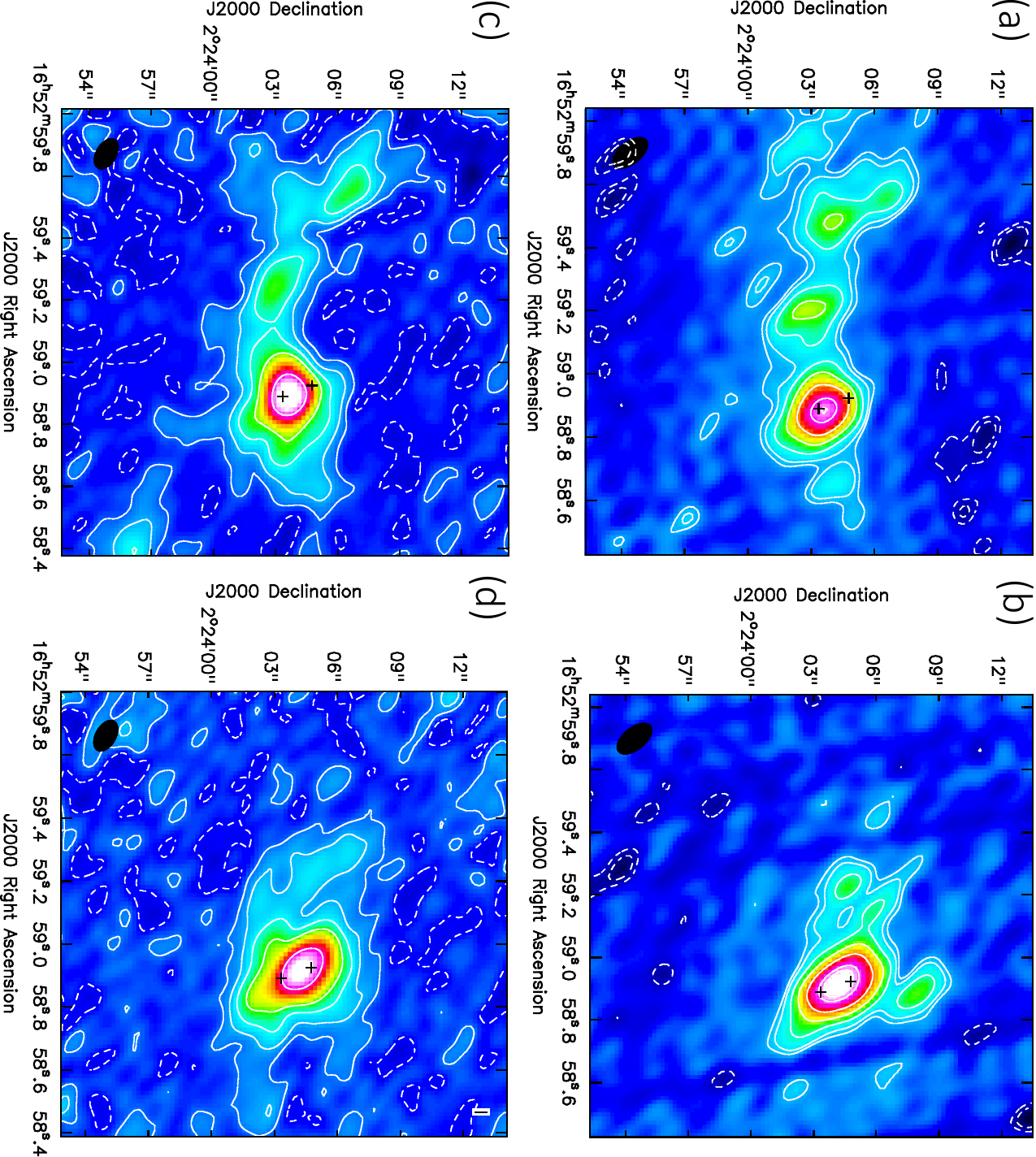}
\caption{\co~({\it top row}) and \cotwo~({\it bottom row}) interferometric maps of the outflow emission, integrated in the blue {\it(a, c)} and red {\it(b, d)} wings, by using the same velocity ranges as in Fig~\ref{fig:outflow_maps}. The maps have matched spatial resolution ($\sim1.2\arcsec$, details in $\S$~\ref{sec:obs}). 
Contours correspond to: ($-3\sigma$, $-2\sigma$, $2\sigma$, $3\sigma$, $6\sigma$, $12\sigma$, $24\sigma$, $48\sigma$, $60\sigma$) with $1\sigma$= 0.33 mJy~beam$^{-1}$ in panel (a) and $1\sigma$= 0.39 mJy~beam$^{-1}$ in panel (b); 
($-3\sigma$, $3\sigma$, $10\sigma$, $24\sigma$, $48\sigma$, $200\sigma$, $350\sigma$) with $1\sigma$= 0.33 mJy~beam$^{-1}$
in panel (c) and $1\sigma$= 0.3 mJy~beam$^{-1}$ in panel (d).
Similar to Fig.~\ref{fig:outflow_maps}, the black crosses indicate the VLBI positions of the AGNs from \cite{Hagiwara+11}. }
\label{fig:COwings}
\end{figure*}

Interferometric maps of the \co~and \cotwo~high velocity emissions are shown in Fig.~\ref{fig:COwings} separately for the blue ({\it left panels: a, c}) and red ({\it right panels: b, d}) line wings. To produce these maps, the \co~and \cotwo~{\it uv} visibilities have been integrated within the same velocity ranges as in Fig.~\ref{fig:outflow_maps}. Figure~\ref{fig:COwings} demonstrates that the blue and red wings of the low-$J$ CO lines in NGC~6240 are both spatially extended on scales of several kpc, with their most extended features aligned preferentially along the east-west direction. The positive contours of the \cotwo~blue and red line wing emissions shown in Fig.~\ref{fig:COwings}(c,d) are overplotted in Figure~\ref{fig:CO21wings_overlap} to allow a direct comparison of their extent and morphology. 

\section{Box spectra and outflow identification}\label{sec:boxspectra_appendix}

\begin{sidewaysfigure}
	\centering
	\includegraphics[width=\textwidth]{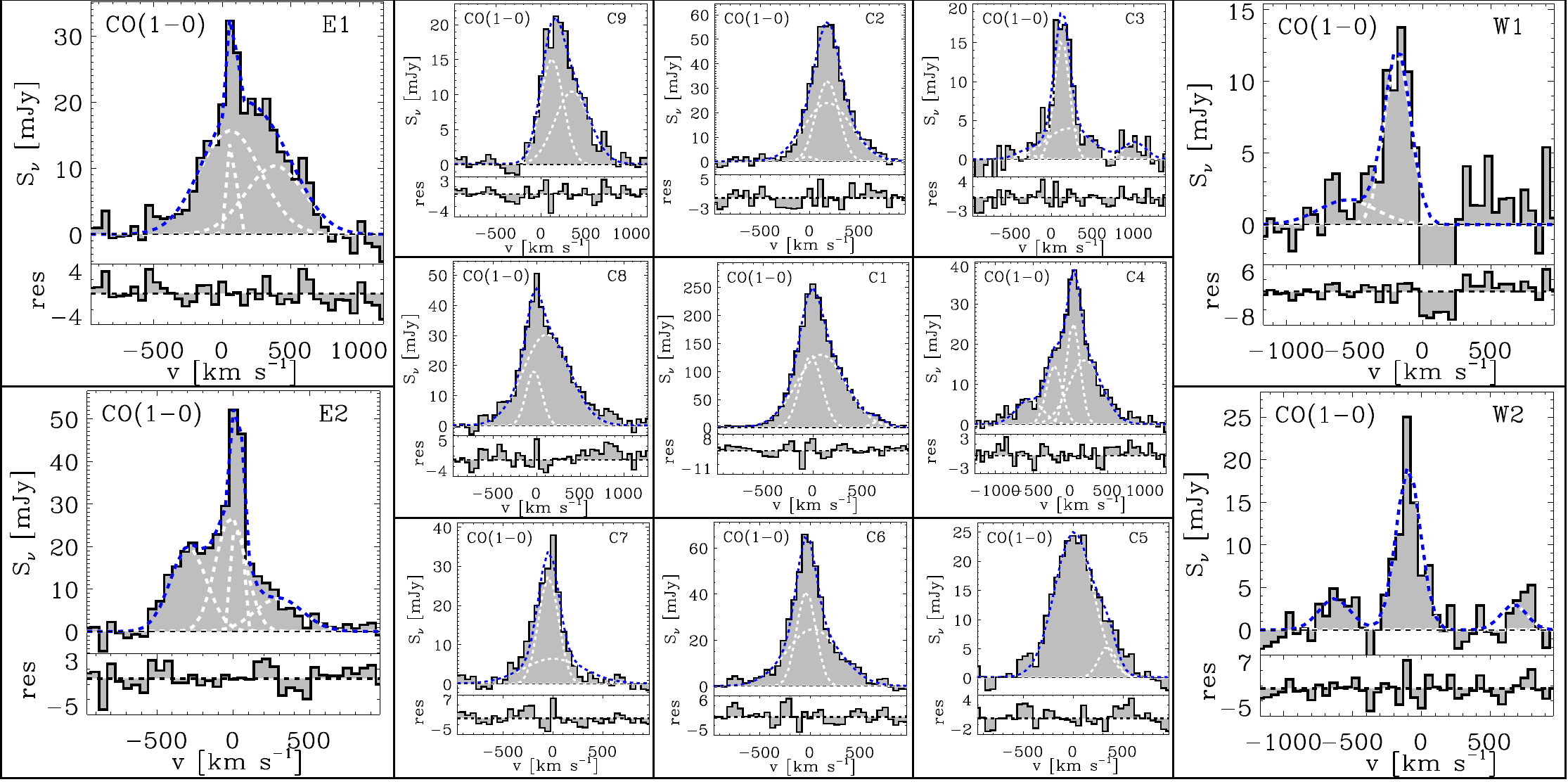}
    \caption{\co~spectra extracted from the grid of 13 boxes shown in panel (a) of Fig~\ref{fig:outflow_maps}, with overplotted the results of the simultaneous fit procedure.}
    \label{fig:boxes_fit_co10}
\end{sidewaysfigure}

\begin{sidewaysfigure}
	\centering
	\includegraphics[width=\textwidth]{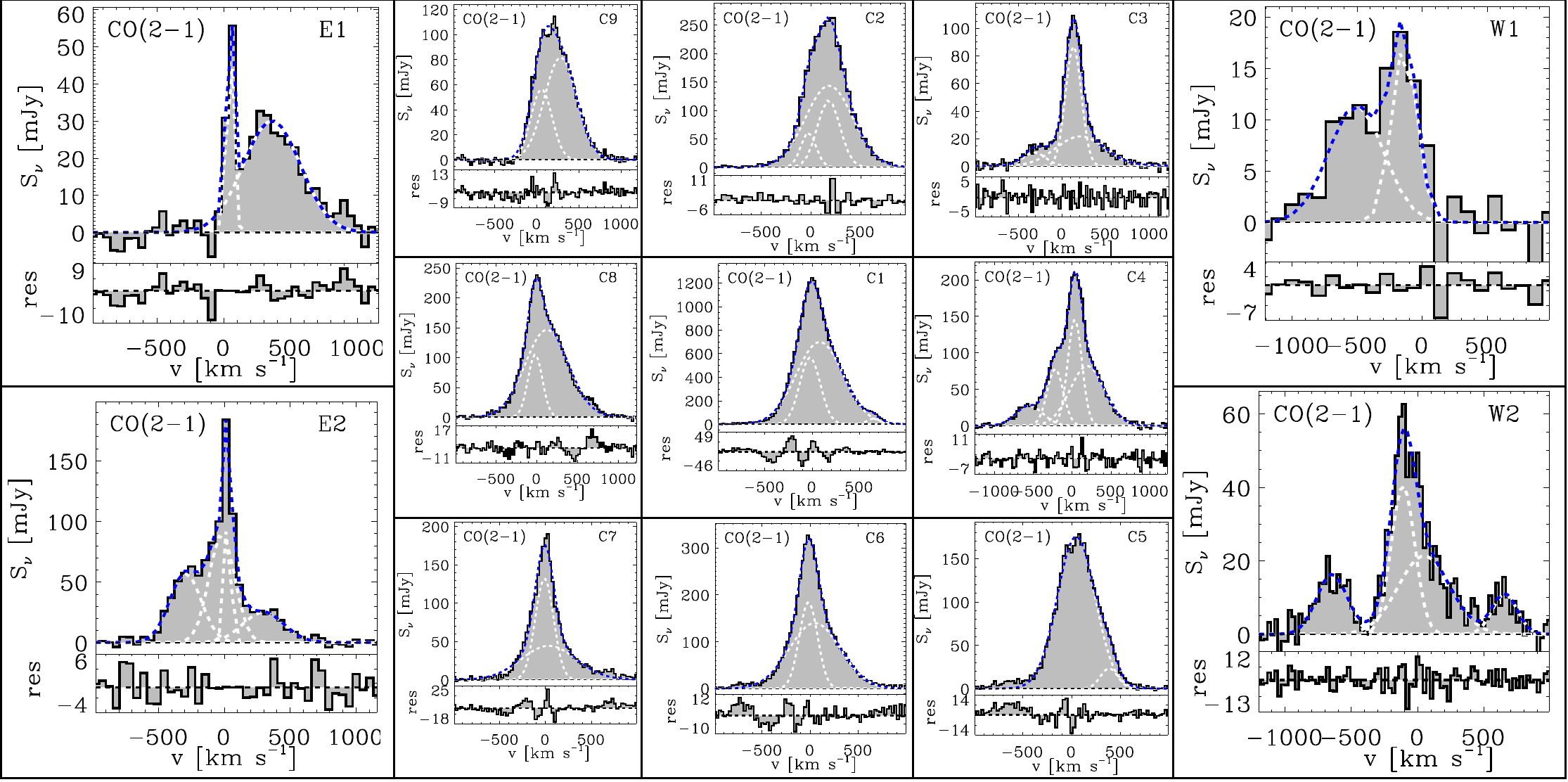}
    \caption{\cotwo~spectra extracted from the grid of 13 boxes shown in panel (a) of Fig~\ref{fig:outflow_maps}, with overplotted the results of the simultaneous fit procedure.}
    \label{fig:boxes_fit_co21}
\end{sidewaysfigure}

\begin{sidewaysfigure}
	\centering
	\includegraphics[width=\textwidth]{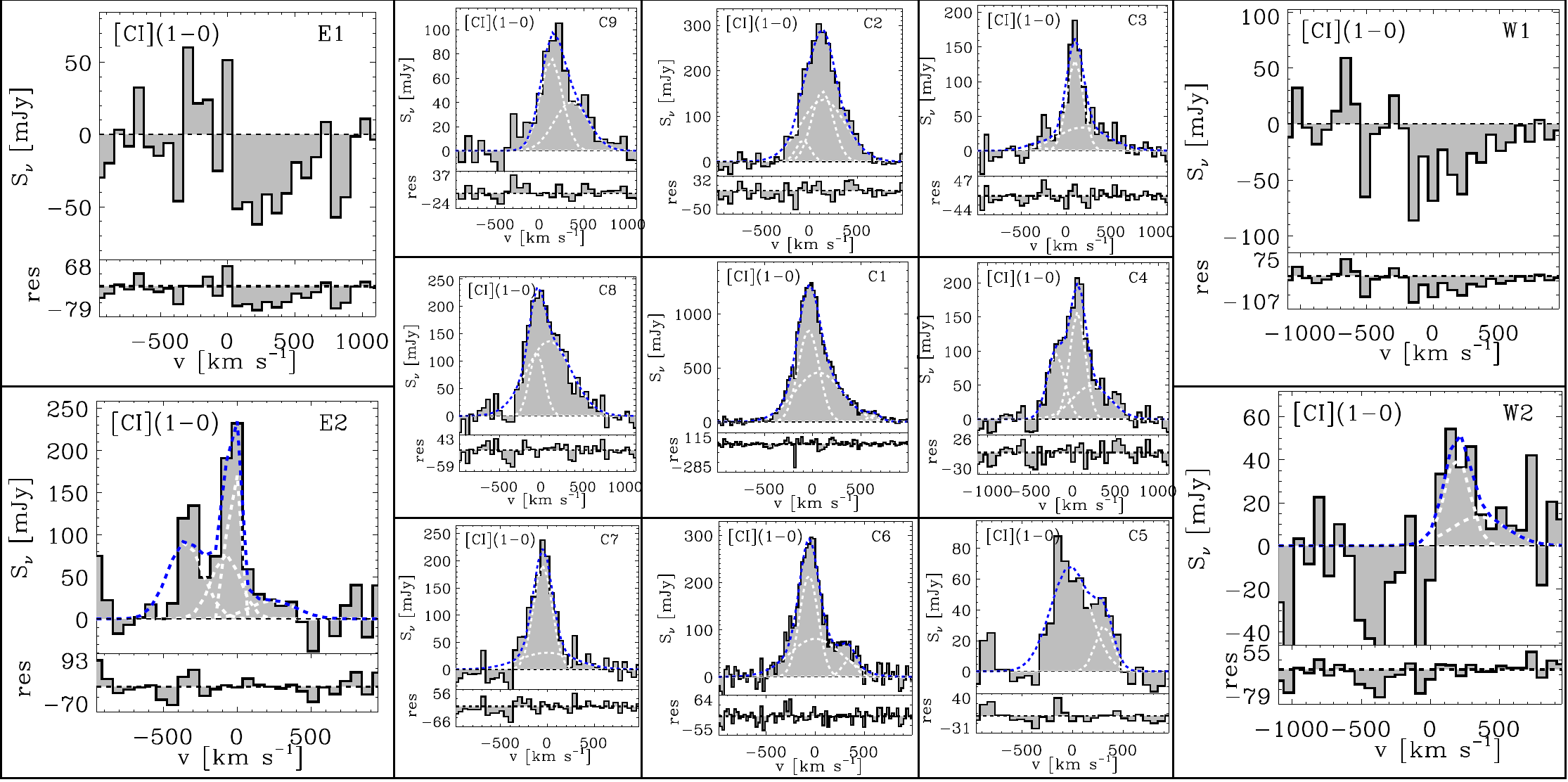}
    \caption{\carbon~spectra extracted from the grid of 13 boxes shown in panel (a) of Fig~\ref{fig:outflow_maps}, with overplotted the results of the simultaneous fit procedure.}
    \label{fig:boxes_fit_ci10}
\end{sidewaysfigure}

The \co, \cotwo, and \carbon~box spectra extracted from the grid shown in panel (a) of Fig~\ref{fig:outflow_maps} are presented in Figures~\ref{fig:boxes_fit_co10}, \ref{fig:boxes_fit_co21}, and~\ref{fig:boxes_fit_ci10}. The results of the simultaneous fitting procedure described in $\S$~\ref{sec:results_boxes} are overplotted on the data. Each spectrum is labelled with the box ID: IDs \texttt{C1-C9} correspond to the central $2\arcsec\times2\arcsec$ boxes, whereas \texttt{E1-E2} and \texttt{W1-W2} are respectively the eastern and western $3\arcsec\times3\arcsec$ boxes.
The spectral components resulting from the simultaneous fit are classified as `outflow' or `quiescent' according to their velocity shift and dispersion, as described in $\S$~\ref{sec:results_boxes}. 
In the following we examine - case by case - the results of such outflow identification procedure (see also $\S$~\ref{sec:disc_caveats_outflowid} for a more general discussion).

It is already evident from Fig.\ref{fig:outflow_maps}(a,b,c) that the bulk of the molecular line emission at high projected velocities 
traces a very extended, non-collimated structure aligned east-west, with a large overlap between the blue and redshifted sides (Figure~\ref{fig:CO21wings_overlap}).
As best seen in the high S/N \cotwo~spectra in Fig.~\ref{fig:boxes_fit_co21}, the boxes \texttt{E1}, \texttt{E2}, \texttt{W1}, and \texttt{W2}, 
tracing gas at $d>2$~kpc from the nucleus, exhibit broad spectral features - distinguishable from the narrow components - 
which in some cases (e.g. \texttt{E1}) dominate the total CO flux. 
Such broad wings are characterised by velocity shifts 
($v\sim300-600$~\kms) and dispersions ($\sigma_{v}\sim125 - 215$~\kms) that are
much higher than the narrow components detected in the same spectra
($v\sim30-180$~\kms, $\sigma_{v}\sim30-100$~\kms).
As a result, we identify the high-$v$, high-$\sigma_v$ spectral components at $d>2$~kpc
as due to an outflow. 

Closer to the nucleus, the outflow identification becomes more challenging. However, once we have identified the broad components
in \texttt{E1}, \texttt{E2}, \texttt{W1}, \texttt{W2} as part of an outflow, then it is natural
to ascribe similar broad components - spatially aligned along the east-west axis - to the same outflow. In particular, 
\texttt{C4} and \texttt{C8} (east and west of the nucleus), \texttt{C2}-\texttt{C3} (north-west
of the nucleus, along the same direction as the high-$v$ structure in Fig.~\ref{fig:outflow_maps}e) and \texttt{C6}-\texttt{C7} (south-east of the nucleus, along another direction of outflow expansion as shown in Fig.~\ref{fig:outflow_maps}d) also show a broad 
component extending up to $\sim1000$~\kms on both the red- and blue-shifted sides. 
The outflow identification is more uncertain for boxes \texttt{C5} and \texttt{C9}, where the wings are less prominent than in other regions, and the spatial alignment with the larger-scale outflow is not obvious. However, the contribution from these boxes to the total outflow mass and mass-loss rate is negligible. Indeed, without \texttt{C5} and \texttt{C9}, we derive
$M_{out}=1.1\times10^{10}$~M$_{\odot}$ and $dM_{out}/dt = 2350$~$M_{\odot}~yr^{-1}$, consistent with the values given in Table~\ref{table:of_prop}. Hence the uncertain outflow identification in boxes \texttt{C5} and \texttt{C9} does not affect our results.

We now discuss the central box \texttt{C1}, which alone contributes to: $M_{out}^{\texttt{C1}} = (4\pm2) \times 10^9~M_{\odot}$ and $dM_{out}^{\texttt{C1}}/dt = 1400\pm1100~M_{\odot}~yr^{-1}$. 
There are several arguments in support of a significant outflow contribution in this region: 
(i) the striking similarity between the \texttt{C1} spectrum with that of the adjacent box \texttt{C8}; 
(ii) the notion that the outflow must originate from within this region, because it hosts two AGNs and most of the star formation activity; 
(iii) the \carbon~emission at $|v|>200$~\kms arising from within this region is spatially extended and 
follows the morphology of the larger-scale outflow (Figure~\ref{fig:outflow_maps}(d,e)); (iv) in this region, the outflow dominates even the emission at
low projected velocities, as shown by Fig.~\ref{fig:CI10_moments}; (iv) the \alphaco~and $r_{21}$ values calculated for the outflow components identified in box \texttt{C1} (data points at $d=0$~kpc in the bottom panels of Fig.~\ref{fig:plots}) are consistent with the values measured in the larger-scale outflow.

Therefore, based on the spectral and spatial properties of the molecular line emission that we have ascribed to the outflow, we conclude that alternative mechanisms such as rotating disks and tidal tails are unlikely to significantly affect our outflow energetics estimates. 

\end{document}